\documentclass[aps,prb,reprint,letterpaper,showpacs,amsmath,amssymb]{revtex4-1}
\usepackage{graphicx}
\usepackage[absolute]{textpos}

\begin{document}
\begin{textblock*}{8.5in}(0.1in,0.25in)
\begin{center}
PHYSICAL REVIEW B \textbf{89}, 245312 (2014)
\end{center}
\end{textblock*}
\begin{textblock*}{2.5in}(5.6in,10.5in)
\copyright2014 American Physical Society
\end{textblock*}
\begin{textblock*}{2.5in}(5.06in,4.31in)
{\small \doi{10.1103/PhysRevB.89.245312}}
\end{textblock*}

\title{Polariton linewidth and the reservoir temperature dynamics in a semiconductor microcavity}
\received{25 April 2014} \revised{12 June 2014} \published{30 June
2014}

\author{V. V. Belykh}
\email[]{belykh@lebedev.ru} \affiliation{Division of Solid State
Physics,\\P.N. Lebedev Physical Institute of the Russian Academy of
Sciences, Leninskii Prospekt 53, Moscow 119991, Russia}

\author{D. N. Sob'yanin}
\affiliation{I.E. Tamm Division of Theoretical Physics,\\P.N.
Lebedev Physical Institute of the Russian Academy of Sciences,
Leninskii Prospekt 53, Moscow 119991, Russia}

\begin{abstract}
A method of determining the temperature of the nonradiative
reservoir in a microcavity exciton-polariton system is developed. A
general relation for the homogeneous polariton linewidth is
theoretically derived and experimentally used in the method. In
experiments with a GaAs microcavity under nonresonant pulsed
excitation, the reservoir temperature dynamics is extracted from the
polariton linewidth. Within the first nanosecond the reservoir
temperature greatly exceeds the lattice temperature and determines
the dynamics of the major processes in the system. It is shown that,
for nonresonant pulsed excitation of GaAs microcavities, the
polariton Bose-Einstein condensation is typically governed by
polariton-phonon scattering, while interparticle scattering leads to
condensate depopulation.
\end{abstract}

\pacs{78.67.Pt, 71.36.+c, 05.30.Jp, 78.47.jd}
\maketitle

\section{Introduction}
\label{Intro} Experimental investigation and practical use of
semiconductor structures in most cases require their nonresonant
excitation. The latter leads to a complex evolution of the
electron-hole (e-h) system, and this evolution involves several
processes \cite{Shah1999}: First, internal thermal equilibrium is
established within charge carriers in a time shorter than 1~ps for
GaAs-based quantum well (QW) structures (considered further in the
present paper) \cite{Knox1986, Knox1988}. When the internal
equilibrium has been established, the e-h system is characterized by
a temperature $T$ greater than the lattice temperature
$T_\text{latt}$. Second, the e-h system cools down due to the
emission of optical (fast stage) and acoustical (slow stage) phonons
\cite{Leo1988, Leo1988a, Yoon1996, Rosenwaks1993}. Both processes
are accompanied by the exciton formation. The characteristic time of
the exciton formation ranges from 10~ps to 1~ns and is determined by
the e-h density (see \cite{Szczytko2004} and references therein).
For the excitation above the QW barriers the whole evolution is
accompanied by the capture of charge carriers to the QWs. For
sufficiently deep QW states the capture is relatively fast, with a
time of $\sim 1$~ps, and is assisted by the emission of optical
phonons \cite{Deveaud1988,Barros1993,Oberli1989,Sotirelis1994}.

The temperature $T$ of the e-h system during its cooldown remains
significantly higher than the lattice temperature $T_\text{latt}$
for several hundreds of picoseconds in the low-temperature
experiments \cite{Yoon1996, Leo1988, Leo1988a, Szczytko2004,
Hoyer2005}. As a result, the dynamics of $T$ determines the exciton
fraction and many important properties of the system. An example is
Bose-Einstein condensation (BEC) of excitons, which is hindered in
bulk semiconductors and the QWs without spatial separation of
electrons and holes. The reason for such a hindrance is
insufficiently fast cooling of excitons compared with their
recombination and inelastic collisions, the latter leading to the
formation of the exciton complexes and e-h liquid \cite{Bagaev2010}.

We are interested in the temperature dynamics of the reservoir, the
e-h system in the QWs embedded in a semiconductor microcavity (MC),
and the effect of this dynamics on the properties of MC exciton
polaritons, mixed exciton-photon states. This system attracts
considerable attention, especially inspired by the achievement of
BEC of polaritons \cite{Kasprzak2006} and a number of intriguing
related phenomena, such as quantized vortices, superfluidity, the
Josephson effect, etc. (see \cite{Sanvitto2012} for a review).  The
dynamics of the reservoir internal temperature after a short-pulse
nonresonant excitation is of primary importance because it
determines the possibility of and conditions for the polariton BEC.
Typically, the internal temperature of the e-h system in bare QW
structures is extracted from the Boltzmann tail in the
photoluminescence (PL) spectra, which originates from the e-h plasma
recombination. However, the recombination is rather weak and
requires for a reasonable analysis a high e-h density \cite{Leo1988}
and high quality of QWs \cite{Szczytko2004}. For the QWs embedded in
a MC, the e-h plasma recombination is even more hindered due to the
strong spectrum modification induced by the MC. In some works the
temperature was extracted by analyzing the lower polariton (LP)
population energy distribution \cite{Deng2003, Kasprzak2006,
Deng2006, Balili2007, Kasprzak2008, Levrat2010, Kammann2012}.
However, the temperature so defined is not the reservoir temperature
but, rather, characterizes the degree of nonequilibrium of the
low-wave-vector part of the polariton system.

In the present paper, we propose and justify a new method of
determining the reservoir temperature from the linewidth of the
lower polariton states. We theoretically derive and experimentally
use a general relation for the LP homogeneous linewidth via the rate
of polariton escape, used to find the reservoir temperature, and the
mean polariton occupation number. The extracted reservoir
temperature in the experiments with nonresonant pulsed excitation of
the GaAs MC decays from about $100$~K at a time of $50-100$~ps after
the excitation pulse and relaxes to the lattice temperature
$T_\text{latt}$ in about $1$~ns. The fact that at long times the
extracted temperature follows $T_\text{latt}$ as $T_\text{latt}$ is
changed proves the validity of our method. We conclude that at the
conditions of the polariton Bose-Einstein condensation in GaAs MC
structures the reservoir temperature greatly exceeds the lattice
temperature. This leads to increasing the BEC threshold and
degrading the coherence properties compared with those for the
reservoir in thermal equilibrium with the lattice. Furthermore, as a
result of the large reservoir temperature, BEC in GaAs MCs under
nonresonant pulsed excitation is typically governed by
polariton-phonon scattering, while scattering of polaritons by
excitons and free charge carriers leads to depopulation of the
condensate.

\section{Theory}
\begin{figure}
\begin{center}
\includegraphics[width=0.9\columnwidth]{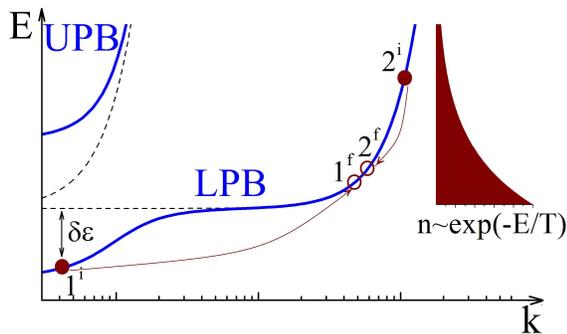}
\caption{Exciton, photon (dashed lines), and polariton (solid lines)
dispersion curves. Arrows show the polariton-exciton scattering
responsible for the LP broadening. Filled area shows the reservoir
energy distribution (population along the horizontal axis and energy
along the vertical axis).} \label{FigRelaxScheme}
\end{center}
\end{figure}

The MC polariton system can be divided into the low-$k$, radiative
part ($k$ is a wave vector) and the high-$k$, nonradiative part,
usually referred to as a reservoir. In the radiative part,
exciton-photon mixing is high, which results in a steep polariton
dispersion curve. This polariton part is strongly nonequilibrium due
to the short polariton lifetime determined by the MC $Q$ factor. On
the other hand, the nonradiative reservoir, containing almost all of
the population of the e-h system, is virtually unaffected by the MC,
which can only alter the rate of the reservoir density decay induced
by exciton scattering to the leaky low-$k$ region \cite{Bajoni2006}.
All the processes occurring in the e-h system and discussed for bare
QW structures in the Introduction also take place in the reservoir.
The reservoir determines the population of the low-$k$ region.

Here, we determine the LP linewidth. Let us consider a subsystem
representing one low-$k$ polariton state in the MC. The evolution of
the probability distribution for the state occupation number $n$ is
governed by the master equation
\begin{equation}
\label{masterEquation} \dot{p}_n=w n p_{n-1}-[w(n+1)+\gamma
n]p_n+\gamma(n+1)p_{n+1},
\end{equation}
where $p_n$ is the probability of finding $n$ polaritons in the
subsystem, with $n=0,1,2,...$, and $w$ and $\gamma$ are the rates
of, respectively, emission and absorption of a polariton by an
environment, the state of which is assumed to be virtually
unaffected by coupling to the subsystem. The stationary solution of
Eq.~\eqref{masterEquation} is
$p^\text{st}_n=(1-w/\gamma)(w/\gamma)^n$ and represents the
Bose-Einstein probability distribution. Let us stress in this
connection that the subsystem is, in general, in a
\emph{nonequilibrium} stationary state and that the environment is
of the general kind and need not be an equilibrium
particle-and-energy bath. The stationary mean polariton number is
\begin{equation}
\label{meanPolaritonNumber} \langle n\rangle=\frac{1}{\gamma/w-1},
\end{equation}
the finiteness of which implies $w<\gamma$.

Now we find the polariton energy spectrum
\begin{equation}
\label{generalSpectrum}
S(E)=\frac{1}{2\pi}\int_{-\infty}^\infty\!\!\!\exp(iE\tau)g^{(1)}(\tau)\,d\tau
\end{equation}
with the normalization $\int S(E)dE=1$, where $g^{(1)}(\tau)=\langle
a^+(0)a(\tau)\rangle/\langle n\rangle$ is the first-order temporal
correlation function (the quantum degree of first-order temporal
coherence) and $a^+$ and $a$ are the creation and annihilation
operators \cite{Loudon2000}. Note that we put $\hbar=k_\text{B}=1$.
Equation~\eqref{generalSpectrum} has the form of the Wiener-Khinchin
theorem \cite{Wiener1930,Khintchine1934,Einstein1914,Yaglom1987}.
Neglecting the interaction between polaritons within the subsystem,
we recover from Eq.~\eqref{masterEquation} the quantum master
equation for the reduced density operator $\rho$ of the subsystem,
\begin{eqnarray}
\dot{\rho}&=&-iE'_0[a^+a,\rho]-\frac{w}{2}(aa^+\rho-2a^+\rho a+\rho aa^+)\nonumber\\
& &-\frac{\gamma}{2}(a^+a\rho-2a\rho a^++\rho
a^+a),\label{quantumMasterEquationForRho}
\end{eqnarray}
where the energy $E'_0$ is close to the energy of the state
uncoupled from the environment. By finding the temporal behavior of
$\langle a\rangle$ from Eq.~\eqref{quantumMasterEquationForRho} and
using the quantum regression theorem
\cite{Lax1963,Lax1967,Carmichael1993}, we have
\begin{equation}
\label{g1Tau} g^{(1)}(\tau)=\exp\biggl(-i
E'_0\,\tau-\frac{\gamma-w}{2}\,|\tau|\biggr).
\end{equation}

We may draw an analogy between the subsystem of noninteracting
polaritons and chaotic light. The analogy stems from the fact that
the probability distribution $\{p^\text{st}_n\}$ is formally similar
to the Planck distribution; in other words, the statistical
properties of the subsystem are similar to those of the chaotic
light emitted by an equilibrium thermal source. From this analogy we
immediately write for the subsystem of polaritons the relation that
takes place for chaotic light \cite{Loudon2000},
\begin{equation}
\label{g2ViaG1} g^{(2)}(\tau)=1+|g^{(1)}(\tau)|^2,
\end{equation}
where $g^{(2)}(\tau)=\langle
a^+(0)a^+(\tau)a(\tau)a(0)\rangle/\langle n\rangle^2$ is the
second-order temporal correlation function (the quantum degree of
second-order temporal coherence).

Interestingly, this analogy allows us to foresee Eq.~\eqref{g1Tau}
directly from Eq.~\eqref{masterEquation} without using
Eq.~\eqref{quantumMasterEquationForRho}: By determining the temporal
behavior of the mean polariton number $\langle n\rangle$ from
Eq.~\eqref{masterEquation} and using the quantum regression theorem,
we arrive at
\begin{equation}
\label{g2} g^{(2)}(\tau)=1+\exp[-(\gamma-w)|\tau|].
\end{equation}
Clearly, we have super-Poissonian fluctuations with $g^{(2)}(0)=2$,
as it must be for the Bose-Einstein distribution. From
Eqs.~\eqref{g2ViaG1} and~\eqref{g2} it follows that
$|g^{(1)}(\tau)|=\exp[-(\gamma-w)|\tau|/2]$, which implies
Eq.~\eqref{g1Tau}.

Finally, from Eqs.~\eqref{generalSpectrum} and \eqref{g1Tau} we
conclude that the polariton spectrum is a Lorentzian
\[
S(E)=\frac{1}{\pi}\frac{\Gamma/2}{(E-E'_0)^2+(\Gamma/2)^2}
\]
with the linewidth (FWHM)
\begin{equation}
\label{polaritonLinewidth} \Gamma=\gamma-w.
\end{equation}
Using Eq.~\eqref{meanPolaritonNumber}, we can also rewrite the
polariton linewidth \eqref{polaritonLinewidth} in an alternative
form
\begin{equation}
\label{polaritonLinewidthViaGammaAndNAv}
\Gamma=\frac{\gamma}{\langle n\rangle+1}.
\end{equation}

Equations \eqref{polaritonLinewidth} and
\eqref{polaritonLinewidthViaGammaAndNAv} are valid for a general
environment with the rates $w$ and~$\gamma$ being arbitrary in
nature. In the particular case of a thermalized exciton reservoir
and the absence of polariton-phonon scattering,
Eq.~\eqref{polaritonLinewidthViaGammaAndNAv} reduces to the known
result \cite{PorrasTejedor2003}.

In our system, the rate of change of the environment state is much
less than $\Gamma$; in other words, the subsystem evolves
adiabatically and all of the above description takes place at each
instant of time. The rate $\gamma$ of polariton escape to the
environment is determined by photon escape through the MC mirrors
with a rate $\gamma_\text{c}$ and polariton scattering assisted by
phonons, excitons, and free carriers (electrons and holes) with the
corresponding rates $\gamma_\text{phon}$, $\gamma_\text{x}$,
$\gamma_\text{e}$, and $\gamma_\text{h}$; thus, $\gamma=
\gamma_\text{c}C^2 + \gamma_\text{phon} + \gamma_\text{x} +
\gamma_\text{e} + \gamma_\text{h}$, where $C$ is the photon Hopfield
coefficient. The rate $\gamma_\text{c}C^2$ is independent of the
reservoir concentration and temperature and hence is time
independent. Under the assumption that the reservoir occupation
numbers are much less than 1, $\gamma_\text{phon}$ is also time
independent.

Let us calculate the rates $\gamma_\text{x}$, $\gamma_\text{e}$, and
$\gamma_\text{h}$. Figure~\ref{FigRelaxScheme} shows the scheme of
polariton-exciton scattering: a polariton scatters off an exciton
and makes a transition from the considered low-$k$ polariton state
$1^\text{i}$ to a reservoir state $1^\text{f}$, with the exciton
making a transition from a state $2^\text{i}$ to a state
$2^\text{f}$. Since the reservoir region contains the overwhelming
majority of the states, it is natural to assume that polaritons
escape mostly to the reservoir due to the scattering off the
reservoir excitons and free carriers. The reservoir is assumed to be
in internal thermal equilibrium \cite{Knox1986, Knox1988, Rota1993,
Alexandrou1995}. Since the e-h density used in the experiment is far
below the saturation density, we have for the reservoir the
Boltzmann distribution with a temperature $T$, which is, in general,
different from the lattice temperature $T_\text{latt}$.

For polariton-exciton scattering we have, according to Fermi's
golden rule,
\begin{eqnarray}
\gamma_\text{x}&=&2\pi %
\iiint
|M|^2 \frac{g_\text{x} 2\pi k_1^\text{f} dk_1^\text{f} \,k_2^\text{f} dk_2^\text{f}\,d\phi\,A^2}{(2\pi)^4} f\boldsymbol{(}E(k_2^\text{i})\boldsymbol{)} %
\nonumber\\
& &\times\delta\boldsymbol{(}E(k_{1}^\text{f})+E(k_{2}^\text{f}) -
E(k_{1}^\text{i})-E(k_{2}^\text{i})\boldsymbol{)}, \label{EqWxGen}
\end{eqnarray}
where integration is performed over the final states of both
particles because the initial state of the first particle is fixed
\emph{a priori} and the initial state of the second particle is
fixed by the momentum conservation law, $g_\text{x}$ is the exciton
spin degeneracy, $\phi$ is the angle between the wave vectors
$\mathbf{k}_{1}^\text{f}$ and $\mathbf{k}_{2}^\text{f}$, $A$ is the
area of the system, $f(E)=(2\pi N_\text{x}/g_\text{x} m_\text{x} T)
\exp(-E/T)$ is the Boltzmann distribution for the exciton gas with a
time-dependent concentration $N_\text{x}$ and temperature $T$, and
$m_\text{x}$ is the exciton effective mass. We choose the bottom of
the exciton dispersion curve as an energy reference point and denote
by $\delta\varepsilon = -E(k_{1}^\text{i})$ the depth of state
$1^\text{i}$ (Fig.~\ref{FigRelaxScheme}). As we consider scattering
from the radiative polariton region with relatively small wave
vectors $k<\omega/c \ll \sqrt{2m_\text{x}\delta\varepsilon}$, where
$\omega$ is the frequency of the light emitted by the MC, we can put
$k_{1}^\text{i}=0$ in the momentum conservation law:
$\mathbf{k}_{2}^\text{i}=\mathbf{k}_{1}^\text{f}+\mathbf{k}_{2}^\text{f}$.
For the matrix element $M$ we take the limit of low momenta and
write \cite{Tassone1999, Porras2002}: $M=X M_\text{x-x}+C
M_\text{sat}$, where $X$ and $C$ are, respectively, the exciton and
photon Hopfield coefficients for state~$1^\text{i}$, $C^2 = 1-X^2 =
(1+\Omega_\text{R}^2/4\delta\varepsilon^2)^{-1}$, with
$\Omega_\text{R}$ being the Rabi splitting; the $M_\text{x-x}$
(exciton-exciton) and $M_\text{sat}$ (saturation) terms describe the
scattering of the exciton and photon components. We neglect the
saturation term and take the matrix element in the form $M=X
E_\text{x-x}a_\text{B}^2/A$, where $a_\text{B}$ is the exciton Bohr
radius and $E_\text{x-x}$ is an effective exciton-exciton
interaction energy constant that considers all possible spin
channels in Eq.~\eqref{EqWxGen}.

Now we obtain from Eq.~\eqref{EqWxGen} an analytical expression for
the escape rate
\begin{equation}
\gamma_\text{x} = \frac{1}{2} m_\text{x} X^2 E_\text{x-x}^2
a_\text{B}^4 N_\text{x}
\exp\biggl(-\frac{2\delta\varepsilon}{T}\biggr). \label{EqWx}
\end{equation}
Similarly, we get an expression for polariton-electron (hole)
scattering:
\begin{eqnarray}
\gamma_\text{e(h)} &=&
\frac{m_\text{e(h)}}{1+m_\text{e(h)}/m_\text{x}} X^2
E_\text{x-e(h)}^2 a_\text{B}^4 N_\text{e(h)}
\nonumber\\
& &\times\exp\biggl(-\frac{(1+m_\text{e(h)}/m_\text{x})
\delta\varepsilon}{T}\biggr), \label{EqWe}
\end{eqnarray}
where $m_\text{e}(m_\text{h})$ and $N_\text{e}(N_\text{h})$ are,
respectively, the electron (hole) effective mass and concentration.

Finally, we can describe the dependence of the polariton linewidth
on time $t$ by the following equation:
\begin{eqnarray}
\label{EqGammaFinal}%
\Gamma(t)&=&\frac{\gamma_0+\delta\gamma(t)}{\langle n\rangle(t)+1}
\nonumber\\
&=&\frac{\gamma_0+r X^2 N(t) \exp[-\alpha \delta\varepsilon /
T(t)]}{\langle n\rangle(t)+1},
\end{eqnarray}
where $\gamma_0=\gamma_\text{c}C^2 + \gamma_\text{phon}$ is the
time-independent rate of polariton escape; $r$ is a constant
determined by the interparticle interaction; the factor
$1<\alpha\leq 2$ depends on the dominant polariton scattering
mechanism, where $\alpha=2$ for polariton-exciton scattering and
$\alpha=1.2 (1.8)$ for polariton-electron (hole) scattering in GaAs
QWs; and $N$ is the density of the reservoir particles by which
polaritons are mostly scattered.

Further, we determine experimentally the polariton linewidth
$\Gamma$ and occupation number $\langle n\rangle$ and use the
described theory to extract from these quantities the polariton
escape rate $\gamma$. The dependence of $\gamma$ on $t$ and
$\delta\varepsilon$ gives a clue to the reservoir temperature
dynamics.

\section{Experimental details}
The sample under study is a $3\lambda/2$ MC with the Bragg
reflectors made of 17 (top mirror) and 20 (bottom mirror) AlAs and
Al$_{0.13}$Ga$_{0.87}$As pairs and providing a $Q$ factor of about
$2000$. Two stacks of three tunnel-isolated In$_{0.06}$Ga$_{0.94}$As
QWs are embedded in the GaAs cavity at the positions of the two
electric-field antinodes of the MC. The Rabi splitting of the sample
is $\Omega_\text{R} \approx 6$~meV. The same sample was used in the
Refs. \cite{Belykh2011,Belykh2012}. The experiments are done at the
photon-exciton detunings $\Delta = -0.2$ and $+2.0$~meV.

The sample is mounted in a He-vapor optical cryostat and excited by
the emission of a mode-locked Ti:sapphire laser generating a
periodic train of 2.5-ps-long pulses at a repetition rate of 76~MHz.
The excitation laser beam is focused into a 120-$\mu$m spot on the
sample surface using a miniature 8~mm focus lens with the optical
axis inclined by $60^{0}$ with respect to the sample normal. In the
nonresonant excitation experiments, the exciting photon energy of
1.596~eV is above the MC mirrors stop band and larger than the GaAs
bandgap. In the experiments with resonant excitation of the LP
branch, the exciting photon energy of $1.4585 \pm 0.0003$~eV is near
the energy of a bare exciton (note the $60^0$ excitation). The
excitation power $P$ is measured before the laser beam has entered
the cryostat, so the presented values of $P$ do not take into
account the transmission of the cryostat windows and focusing lens,
which lower $P$ by about $30\%$. The PL is collected by a 6-mm focus
micro-objective located in front of the sample surface so that the
surface is near its focal plane. Both the focusing lens and the
micro-objective are mounted on the sample holder inside the
cryostat, and this provides good stability of the system against
vibrations. The PL coming out from the cryostat is focused with a
76-mm focus lens to form an intermediate magnified image of the PL
spot. A 0.7-mm-diameter diaphragm is inserted in the image plane and
selects a 60-$\mu$m-diameter region of the spot with a homogeneous
PL intensity distribution. Then the selected PL passes through a
30-mm lens to fall on the slit of a spectrometer coupled to a
Hamamatsu streak camera. The spectrometer slit is located in the
focal plane of the lens. Thus, the emission angle of the PL is
transformed into the spatial coordinate and selected by the
spectrometer and streak camera slits, which provides a resolution of
about $1^0$. By moving the final lens, it is possible to change the
selected angle. The time and spectral resolutions of this system are
20-30~ps and 0.2-0.3~meV, respectively.

The time-resolved spectra for a given time $t$ after the excitation
pulse are obtained by integrating the emission in the time range
$[t-25, t+25]$~ps for nonresonant excitation experiments and $[t-5,
t+5]$~ps for resonant excitation experiments.

\section{Results and discussion}
\begin{figure*}
\begin{center}
\includegraphics[width=1.75 \columnwidth]{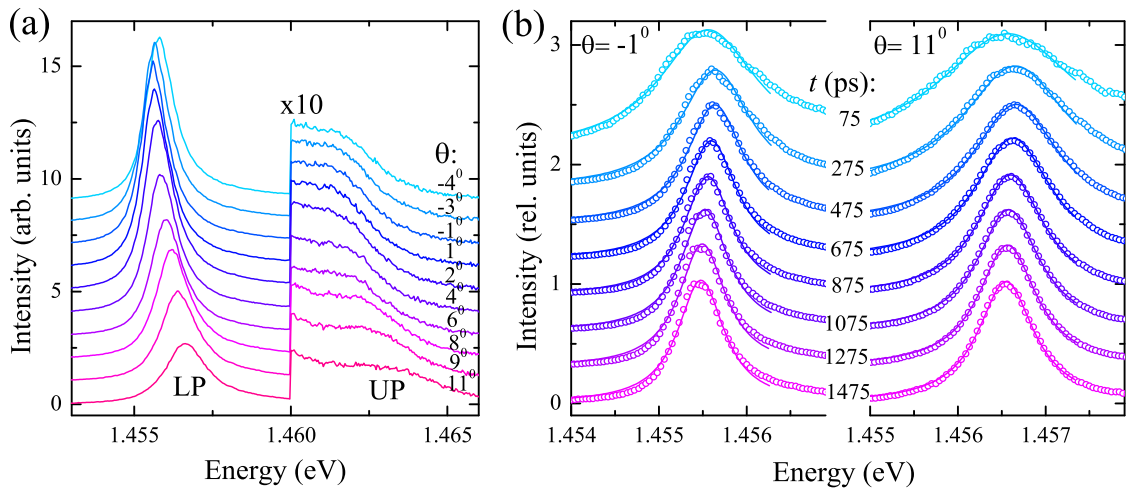}
\caption{(a) MC emission spectra corresponding to different angles
of observation $\Theta$ for $t=275$~ps. Spectra are vertically
shifted and intensity values for $E>1.46$~eV are multiplied by 10
for clarity. (b) LP spectra corresponding to different times after
the excitation pulse (circles) for two angles of observation.
Spectra are normalized to the maximum value and vertically shifted.
Solid lines show Lorentzian fits. In (a) and (b) $\Delta=-0.2$~meV,
$T_\text{latt}=10$~K, nonresonant excitation with $P=1$~mW.}
\label{FigSpectra}
\end{center}
\end{figure*}

\begin{figure}
\begin{center}
\includegraphics[width=\columnwidth]{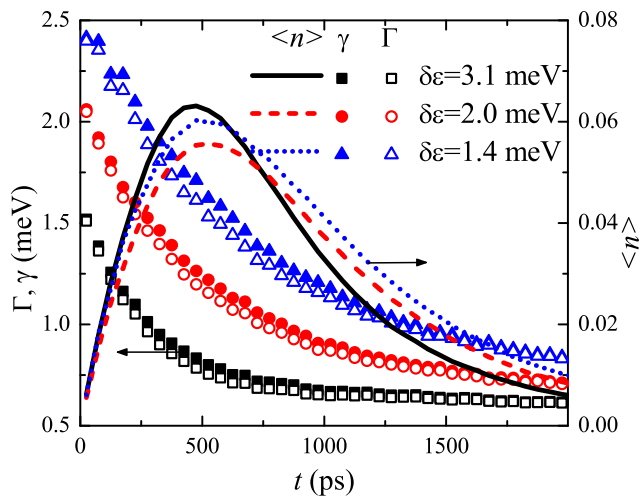}
\caption{Kinetic dependencies of the LP linewidth $\Gamma$ (open
symbols), polariton escape rate $\gamma$ (solid symbols), and
population $\langle n\rangle$ (lines, right axis) for different
state depths $\delta\varepsilon$. The black squares and solid line
correspond to $\Delta = -0.2$~meV, $\Theta = -1^0$, the red circles
and dashed line correspond to $\Delta = -0.2$~meV, $\Theta = 11^0$,
and the blue triangles and dotted line correspond to $\Delta =
2.0$~meV, $\Theta = 13^0$. Nonresonant excitation with $P=1$~mW,
$T_\text{latt}=10$~K.} \label{FigLWKin}
\end{center}
\end{figure}

\begin{figure*}
\begin{center}
\includegraphics[width=1.75\columnwidth]{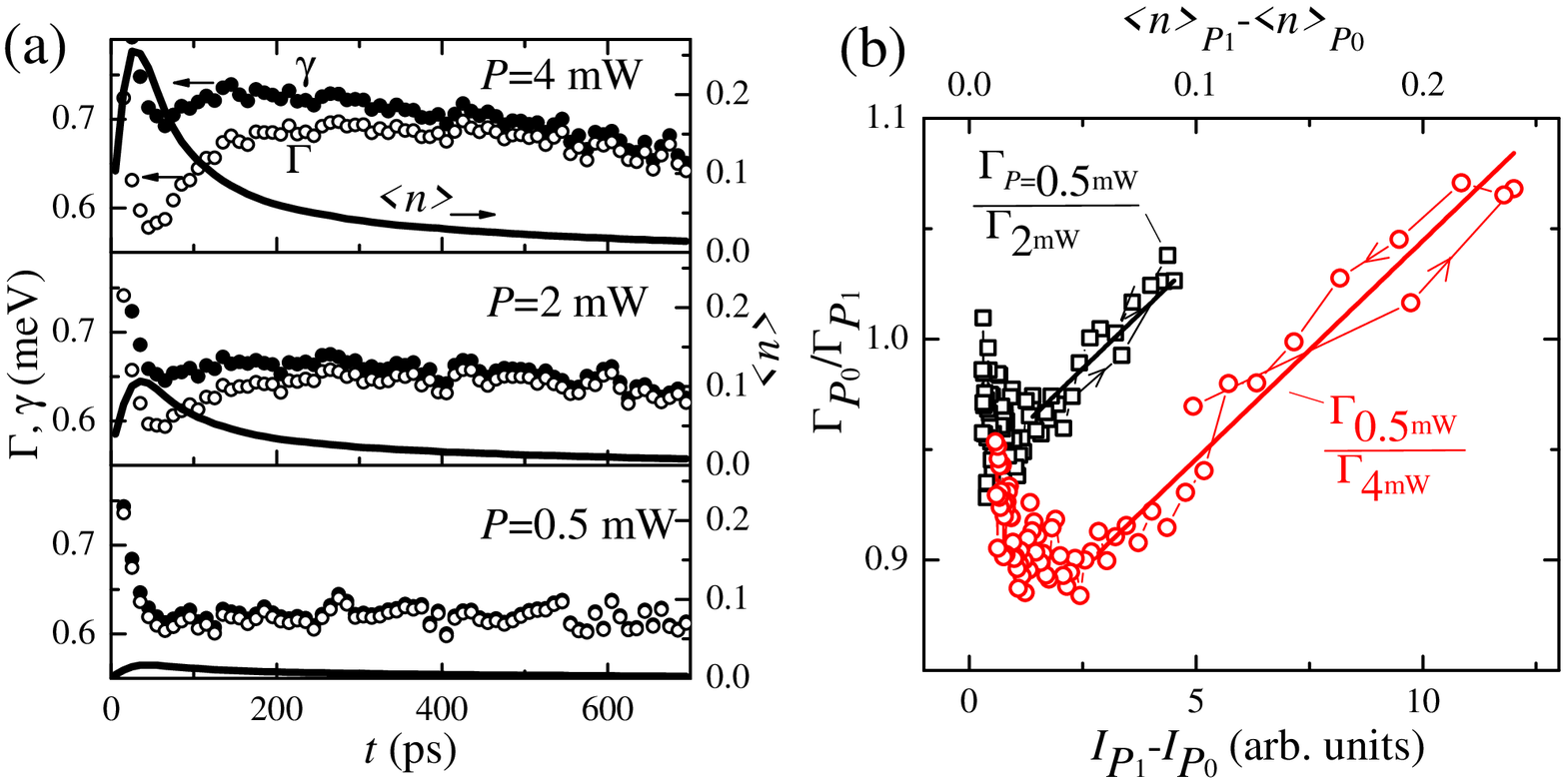}
\caption{Resonant excitation case. (a) Kinetic dependencies of the
LP linewidth $\Gamma$ (open symbols), polariton escape rate $\gamma$
(solid symbols), and population $\langle n\rangle$ (lines, right
axis) for different excitation powers $P$. (b) Ratios of the LP
linewidths for low, $P_0$, and high, $P_1$, excitation powers
$\Gamma(P_0)/\Gamma(P_1)$ as a function of the difference of the
intensities $I(P_1)-I(P_0)$ (bottom axis) and the difference of the
filling factors $\langle n\rangle(P_1)-\langle n\rangle(P_0)$ (top
axis). Data are presented for $P_0=0.5$~mW and two different $P_1$,
$P_1=2$~mW (squares) and $4$~mW (circles). Arrows on the lines
connecting data points indicate the direction of increasing time.
Thich solid lines show linear fits. In (a) and (b)
$\delta\varepsilon =2.1$~meV ($\Delta = 2.0$~meV, $\Theta = 0^0$),
$T_\text{latt}=10$~K.} \label{FigRes}
\end{center}
\end{figure*}

The MC emission spectra corresponding to different angles of
observation (polariton wave vectors) at time $t=275$~ps after the
nonresonant excitation pulse are presented in
Fig.~\ref{FigSpectra}(a) for the photon-exciton detuning $\Delta =
-0.2$~meV. The time-average excitation power $P=1$~mW corresponds to
the electron-hole pair density per QW below $5 \times
10^{10}$~cm$^{-2}$. The spectra show two lines corresponding to the
LP and upper polariton (UP) branches, with the characteristic
angular dependencies of their energies indicating strong
exciton-photon coupling.  The measured LP linewidth (FWHM) $\Gamma$
is mainly determined by the rates of polariton scattering and photon
escape, the processes giving a Lorentzian intensity distribution, as
discussed above. The measurements of the highly photon-like LP
linewidth give $\gamma_\text{c} \approx 1$~meV. Thus, for $\Delta =
-0.2$~meV and $\Theta = -1^0$ the contribution of photon escape to
the linewidth is $\gamma_\text{c} C^2 \approx 0.5$~meV.
Inhomogeneous broadening, mainly related to the QW width
fluctuations, and the instrumental response also give some
contribution to $\Gamma$ in the form of a Gaussian component. The
best fit to the LP line for $\Delta = -0.2$~meV and $\Theta = -1^0$
at long $t$ with the Voigt function gives a Lorentzian component
width of $\approx 0.5$~meV and a Gaussian component width of
$\approx 0.3$~meV, close to $\gamma_\text{c} C^2$ and the
instrumental response function width, respectively.  On the other
hand, the best fit to the same spectrum with the Lorentzian
distribution gives $\Gamma \approx 0.6$~meV. Thus, the relative
contribution of the nonhomogeneous sources is small for the
considered photon-exciton detunings and observation angles
(corresponding to $\delta\varepsilon = 1.4-3.1$~meV), especially at
shorter times, and further the LP line is fitted by the Lorentzian
distribution to determine the FWHM [Fig.~\ref{FigSpectra}(b)].

The LP line, broad at short times after the nonresonant excitation
pulse, significantly narrows at longer times
[Fig.~\ref{FigSpectra}(b)]. The narrowing rate varies for different
angles of observation, as first pointed out in Ref.
\cite{Belykh2012}. For the small angle $\Theta=-1^0$
($\delta\varepsilon=3.1$~meV) the linewidth is close to its
low-density limit already at $t=475$~ps, while for the larger angle
$\Theta=11^0$ ($\delta\varepsilon=2.0$~meV) the line continues
narrowing for significantly longer times. The LP linewidth $\Gamma$
dynamics for these angles is presented in Fig.~\ref{FigLWKin} by
open squares and circles. However, as follows from the theory
[Eq.~(\ref{EqGammaFinal})], not $\Theta$ but the state depth
$\delta\varepsilon$ is the proper parameter that determines the
linewidth dynamics. Indeed, a decrease in $\delta\varepsilon$
achieved by increasing the photon-exciton detuning $\Delta$ leads to
the further slowdown of the linewidth kinetics, as shown by open
triangles for $\Delta=2.0$~meV and $\Theta=13^0$
($\delta\varepsilon=1.4$~meV). Such behavior of the linewidth
kinetics with $\delta\varepsilon$ can be understood from
Eq.~(\ref{EqGammaFinal}). With increasing $\delta\varepsilon$ the
linewidth becomes more sensitive to the reservoir temperature
dynamics and decreases much faster with decreasing~$T$.

\begin{figure*}
\begin{center}
\includegraphics[width=1.75\columnwidth]{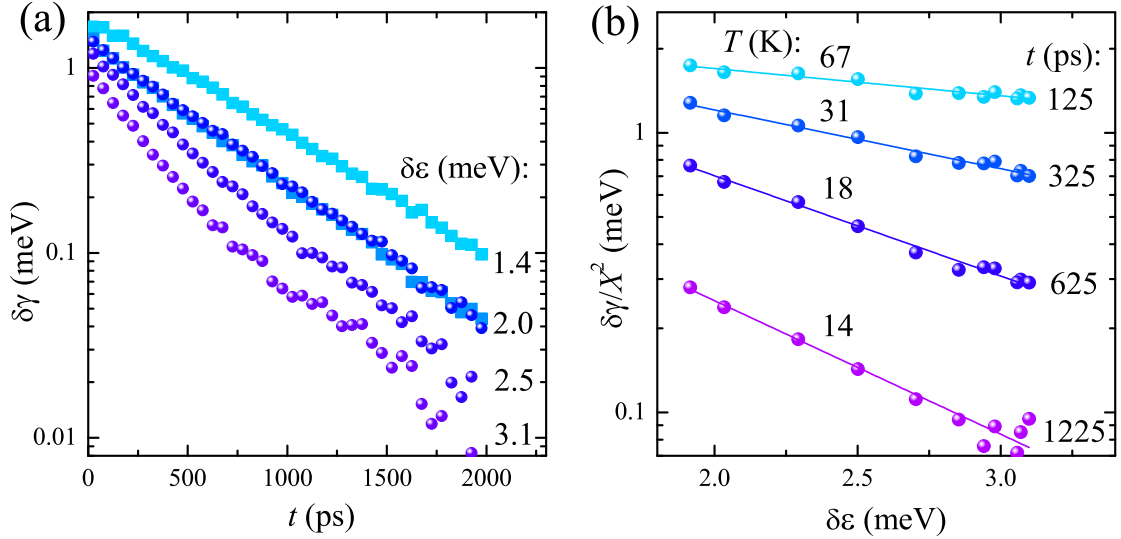}
\caption{(a) Kinetic dependencies of the time-dependent part $\delta
\gamma$ of the polariton escape rate for different state depths
$\delta\varepsilon$ corresponding to $\Delta = -0.2$~meV, $\Theta =
-1^0, 8^0, 11^0$ (circles) and $\Delta = 2.0$~meV, $\Theta = 5^0,
13^0$ (squares). (b) Energy distributions of $\delta \gamma$
normalized to the exciton fraction for different times after the
excitation pulse. Solid lines are exponential fits. $\Delta =
-0.2$~meV. In (a) and (b) $T_\text{latt}=10$~K, nonresonant
excitation with $P=1$~mW.} \label{FiggDistrib}
\end{center}
\end{figure*}

Now we aim to determine the polariton escape rate $\gamma$ from the
measured linewidth $\Gamma$. According to
Eq.~(\ref{polaritonLinewidthViaGammaAndNAv}), $\gamma=\Gamma
(\langle n\rangle+1)$. The mean number $\langle n\rangle$ of
polaritons in a single quantum state is proportional to the emission
intensity $I$:
\begin{equation}
I = \kappa C^2 \langle n\rangle, \label{EqITon}
\end{equation}
where $\kappa = \gamma_\text{c} \times 2 \Delta k_\text{x} \Delta
k_\text{y} A / (2\pi)^2 \times \omega \times \tilde{\kappa}$ is the
photon escape rate times the number of the states from which the
intensity is registered times the energy of emitted photons
$\omega$, and times a constant $\tilde{\kappa}$ that transforms the
real intensity in watts to the intensity measured by the streak
camera in arbitrary units; $\Delta k_\text{x}$ and $\Delta
k_\text{y}$ are the wave-vector intervals determined by the angular
aperture in which the emission is registered. The coefficient
$\kappa$ is independent of $\delta\varepsilon$ and constant in all
the considered experiments (we neglect the small variation of
$\omega$). To determine $\kappa$, we implement the conditions under
which the reduction of $\Gamma$ due to a finite $\langle n\rangle$
is detected directly. For resonant excitation of the LP branch the
reservoir is not overheated, in contrast to the nonresonant
excitation case, and its temperature is close to the lattice
temperature already at short $t$. Thus, the time-dependent
contribution to the escape rate $\gamma$~[Eq.~(\ref{EqGammaFinal})]
is minimum, and the kinetics of the linewidth $\Gamma$ is dominated
by the kinetics of the polariton number $\langle n \rangle$
[denominator in Eq.~(\ref{EqGammaFinal})].

The measured kinetic dependencies of the linewidth for resonant
excitation of the PL branch at $\Theta \approx 60^0$ with different
powers $P$ are presented in Fig.~\ref{FigRes} by open symbols. At
short times, $\Gamma$ experiences a pronounced drop down
proportional to the measured intensity (shown by solid lines, right
axis). The intensity has already been transformed to $\langle n
\rangle$ by dividing by a constant $\kappa C^2$ that will be
determined further. An increase in $P$ leads to an enhancement of
the intensity and the corresponding increase in the $\Gamma$ drop
down. According to Eqs.~(\ref{EqGammaFinal}) and~(\ref{EqITon}),
$\Gamma(t)^{-1} \approx \gamma_0^{-1}[1 - \delta\gamma(t) / \gamma_0
+ \langle n \rangle(t)]=\gamma_0^{-1}[1 - \delta\gamma(t) / \gamma_0
+ I(t)/\kappa C^2]$ for $\delta\gamma(t) / \gamma_0 \ll 1$ and
$\langle n \rangle(t) \ll 1$. The linewidth ratio for two excitation
powers $P_0$ (small) and $P_1$ (high),
\begin{equation}
\frac{\Gamma_{P_0}(t)}{\Gamma_{P_1}(t)} \approx 1 -
\frac{\delta\gamma_{P_1}(t)-\delta\gamma_{P_0}(t)}{\gamma_0} +
\frac{I_{P_1}(t)-I_{P_0}(t)}{\kappa C^2} \label{EqGRatio}
\end{equation}
allows us to eliminate the systematic error, the line broadening at
$t<40$~ps due to the scattered light from higher states. This ratio
is proportional to the intensity difference $I_{P_1}(t)-I_{P_0}(t)$
with the desired coefficient $1/\kappa C^2$ in the time range where
$I(t)$ varies  with $t$ much faster than $\delta\gamma(t)$
($t\lesssim200$~ps). Figure~\ref{FigRes}(b) shows the dependencies
of $\Gamma_{P_0}/\Gamma_{\text{P}_1}$ on $I_{P_1}-I_{P_0}$ for two
different values of $P_1$, and the direction of increasing time is
indicated by arrows. The dependencies are close to linear for high
intensities (at $t\lesssim200$~ps), as expected from
Eq.~(\ref{EqGRatio}). As the intensity first increases and then
decreases with time, the hysteresis in $\Gamma_{P_0}/\Gamma_{P_1}$
is small. This validates the fact that, in the considered time
range, $\delta\gamma(t)$ varies much slower than $I(t)$, and hence,
the slope of the linear dependence of $\Gamma_{P_0}/\Gamma_{P_1}$ on
$I_{P_1}-I_{P_0}$ gives the sought-for coefficient $\kappa C^2$.
From the linear fits to both the dependencies (thick solid lines) we
find $\kappa C^2 = 50 \pm 2$. From the measured intensities we
calculate with Eq.~(\ref{EqITon}) and the known $C(\varepsilon)^2$
the polariton population for all considered states
[Figs.~\ref{FigLWKin} and \ref{FigRes}(b), right axis]. We note that
our method to determine $\langle n \rangle$ is more precise than the
direct method based on measuring the emitted intensity
\cite{Renucci2005} because the latter method requires knowledge of
the exact number of the registered states, which is hard to
determine.

Once $\Gamma$ and $\langle n \rangle$ are found, we calculate with
Eq.~(\ref{polaritonLinewidthViaGammaAndNAv}) the polariton escape
rate $\gamma$, which is shown by solid symbols in
Fig.~\ref{FigLWKin} for nonresonant excitation and in
Fig.~\ref{FigRes}(a) for resonant excitation. It is interesting that
the time-dependent components of both $\Gamma$ and $\gamma$ for
nonresonant excitation are much larger than the corresponding
components for resonant excitation for comparable polariton
populations. This fact is a good illustration of the reservoir
overheating induced by nonresonant excitation and causing the LP
line broadening.

\begin{figure}
\begin{center}
\includegraphics[width= \columnwidth]{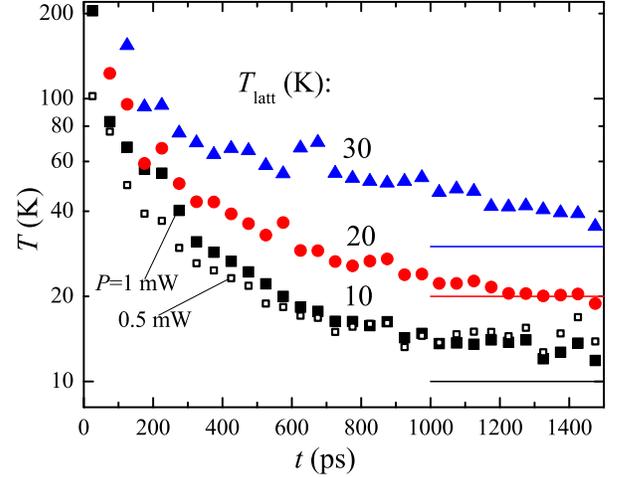}
\caption{Dynamics of the e-h reservoir temperature for different
lattice temperatures $T_\text{latt}$ after nonresonant excitation
with $P=1$~mW (solid symbols). Open squares show the corresponding
dependence for $P=0.5$~mW and $T_\text{latt}=10$~K. $\Delta =
-0.2$~meV. Horizontal lines indicate the values of $T_\text{latt}$.}
\label{FigTDyn}
\end{center}
\end{figure}

Figure~\ref{FiggDistrib}(a) shows the kinetics of the time-dependent
component of the polariton escape rate, $\delta\gamma(t)$, which is
determined as the difference of $\gamma(t)$ (solid symbols in
Fig.~\ref{FigLWKin}) and its value at long $t$. The data are shown
for different state depths $\delta\varepsilon$, and changing
$\delta\varepsilon$ is performed by increasing the observation angle
and photon-exciton detuning [squares and circles in
Fig.~\ref{FiggDistrib}(a) correspond to two different detunings].
For $\delta\varepsilon=2.0$~meV the dependencies corresponding to
different detunings almost coincide despite different angles, which
confirms that $\delta\varepsilon$ is the proper parameter to define
the properties of a polariton state. As $\delta\varepsilon$ is
increased, the kinetics of $\delta\gamma$ becomes faster at $t
\lesssim 1000$~ps. At longer times the decay of $\delta\gamma$ is
more $\delta\varepsilon$ independent. According to
Eq.~(\ref{EqGammaFinal}),
\begin{equation}
\label{EqDgamma}%
\delta\gamma(t, \delta\varepsilon) \propto [X(\delta\varepsilon)]^2
N(t) \exp\biggl(-\frac{\alpha \delta\varepsilon}{T(t)}\biggr),
\end{equation}
and the observed behavior of $\delta\gamma$ indicates a strong
variation of the reservoir temperature $T$ at $t \lesssim 1000$~ps.
The variation becomes smaller at longer times as $T$ relaxes to
$T_\text{latt}$. To make this description more quantitative, we plot
in Fig.~\ref{FiggDistrib}(b) $\delta\gamma/X^2$ as a function of
$\delta\varepsilon$ for different times and fixed $\Delta=-0.2$~meV.
The dependencies are well described by the exponential function, in
accordance with Eq.~(\ref{EqDgamma}). This allows us to determine
the reservoir temperatures $T$ [already indicated in
Fig.~\ref{FiggDistrib}(b)] provided the coefficient $\alpha$ is
known. We obtain $\alpha= 1.3$ from the condition that $T$
approaches $T_\text{latt}$ at long times when $T_\text{latt}=20$~K
(red solid circles in Fig.~\ref{FigTDyn}), and the same value
$\alpha= 1.3$ is fixed for all the considered $T_\text{latt}$ and
excitation powers.

The time dependencies of the reservoir temperature $T$ for different
lattice temperatures $T_\text{latt}$ are presented in
Fig.~\ref{FigTDyn}. At very short times, $T$ is too large to
determine; in the time range $t\sim 50-1000$~ps, $T$ changes from
about 200~K to the values close to $T_\text{latt}$. The fact that at
long $t$ the determined reservoir temperature follows
$T_\text{latt}$ for \emph{different} values of $T_\text{latt}$ is a
proof of the validity of our method. Interestingly, for small
lattice temperatures, $T$ stays considerably larger than
$T_\text{latt}$ for several hundreds of picoseconds, in agreement
with Refs. \cite{Yoon1996, Leo1988, Leo1988a, Szczytko2004,
Hoyer2005}. For the excitation power $P=0.5$~mW (open squares) the
reservoir temperature $T$ at $t \lesssim 1000$~ps is reduced
compared with $T$ for $P=1$~mW (solid squares). This observation
indicates that the reservoir cooldown is slower for an increased
number of particles and can be explained by reabsorption of emitted
phonons (hot-phonon bottleneck effect), which is more effective for
a denser system \cite{Leo1988, Leo1988a, Rosenwaks1993}. Thus, the
obtained dynamics of $T$ is very similar to the e-h temperature
dynamics reported for bare QW structures \cite{Yoon1996, Leo1988,
Leo1988a, Szczytko2004, Hoyer2005}.

It is instructive to discuss the reservoir temperature dynamics
(Fig.~\ref{FigTDyn}) in relation to the MC polariton Bose-Einstein
condensation. In experiments with GaAs MCs under nonresonant pulsed
excitation, BEC is usually observed in the time range $t \lesssim
200$~ps \cite{Tempel2012, Belykh2013}. Our results indicate that in
this time range the reservoir is strongly overheated, calling into
question the existence of excitons in the BEC regime (however, not
canceling the polariton picture \cite{Houdre1994, Tsintzos2008}) and
resulting in the significantly increased BEC threshold and degraded
BEC coherence properties in comparison with what one would expect
for the reservoir in equilibrium with the lattice. Furthermore,
increasing the excitation density well above the threshold leads, on
the one hand, to shortening the BEC onset time \cite{Belykh2013}
and, on the other hand, to increasing the reservoir temperature for
any given time (open and solid squares in Fig.~\ref{FigTDyn}). These
effects can be one of the reasons leading to the suppression of the
condensate spatial coherence for the excitation densities well above
the threshold \cite{Belykh2013}. It is illustrative that, for the MC
structure considered here, further increasing the nonresonant
excitation power leads to lasing on the energy of a photon mode
\cite{Belykh2012}. By contrast, in a similar structure under
resonant excitation, and hence with a cold reservoir, the polariton
BEC was reported \cite{Krizhanovskii2007}.

An interesting conclusion can immediately be drawn from $T\gg
T_\text{latt}$ at BEC. According to Eq.~(\ref{meanPolaritonNumber}),
above the BEC threshold $\gamma\approx w$, which can be rewritten as
\begin{equation}
(w_\text{xeh}- \gamma_\text{xeh}) + (w_\text{phon} - \gamma_\text{phon})-\gamma_\text{c}C^2\approx0,%
\label{EqgEqwMod}
\end{equation}
where $\gamma_\text{xeh} =
\gamma_\text{x}+\gamma_\text{e}+\gamma_\text{h}$ is the rate of the
polariton escape assisted by interparticle interaction
($\gamma_\text{xeh}$ coincides with the time-dependent escape rate
$\delta\gamma$) and $w_\text{phon}$ and $w_\text{xeh}$ are the rates
of the polariton scattering assisted by phonons and interparticle
interaction to the given state. Since the general expression
(\ref{meanPolaritonNumber}) for the mean polariton number reduces to
the Bose-Einstein distribution $\{\exp[(E-\mu)/T]-1\}^{-1}$ when the
state interacts solely with the equilibrium reservoir (which
formally corresponds to $\gamma_\text{c}C^2 = \gamma_\text{phon} =
w_\text{phon} = 0$), we have
\begin{eqnarray}
w_\text{xeh} &=&
\gamma_\text{xeh}\exp\biggl(-\frac{E-\mu}{T}\biggr),
\label{EqDetailEq}
\end{eqnarray}
where $\mu$ is the chemical potential for the exciton part of the
reservoir. Alternatively, this equation can be derived by
calculating $\gamma_\text{xeh}$ and $w_\text{xeh}$ in the Born
approximation. It is easy to show that for a nondegenerate reservoir
\[
\exp\biggl(-\frac{E-\mu}{T}\biggr) \approx \frac{2\pi
N_\text{x}}{g_\text{x} m_\text{x}T}
\exp\biggl(\frac{\delta\varepsilon}{T}\biggr).
\]
Taking the realistic parameters for a GaAs structure
$\delta\varepsilon=5$~meV, $m_\text{x}=0.3 m_0$ (two-dimensional
exciton effective mass \cite{Hillmer1989}), where $m_0$ is the free
electron mass, $g_\text{x}=4$, and taking the reservoir temperature
$T=60$~K at $t \sim 100$~ps (Fig.~\ref{FigTDyn}), we obtain
$\exp[-(E-\mu)/T] \approx N_\text{x} \times (2 \times
10^{-12}$~cm$^2) < 1$ because the polariton BEC implies the strong
coupling regime and, thus, the unsaturated reservoir, $a_\text{B}^2
N_\text{x} \ll 1$, where the exciton Bohr radius is $a_\text{B}\sim
10^{-6}$~cm. With Eq.~(\ref{EqDetailEq}), we conclude that the rate
of polariton scattering to the given state due to interparticle
interaction is smaller than the corresponding polariton escape rate:
\begin{equation}
w_\text{xeh} < \gamma_\text{xeh}. \label{EqWlessG}
\end{equation}
As a result, the first term in Eq.~(\ref{EqgEqwMod}) is negative;
therefore, polariton escape in the regime of BEC is compensated only
by the phonon-assisted polariton relaxation. Thus, contrary to the
common belief, for nonresonant excitation, interparticle interaction
drives polaritons away from the condensate rather than promoting
their condensation.

Condition (\ref{EqWlessG}) might be violated for too deep states
[note $\exp[-(E-\mu)/T] \propto \exp(\delta\varepsilon/T)$], e.g.,
for $\delta\varepsilon > 13$~meV at $N_\text{x}=10^{11}$~cm$^{-2}$
or $\delta\varepsilon > 25$~meV at $N_\text{x}=10^{10}$~cm$^{-2}$.
These values of $\delta\varepsilon$ are relatively large for BEC in
GaAs MCs \cite{Belykh2013, Tempel2012, Balili2007,
Krizhanovskii2007, Wertz2009, Kammann2012} but are easily achievable
for the MCs based on materials with larger exciton binding energy
and Rabi splitting, such as CdTe \cite{Kasprzak2006, Kasprzak2008,
DelValle2009}, GaN \cite{Christopoulos2007, Levrat2010}, and ZnO
\cite{Li2013}. In this case condition (\ref{EqWlessG}) can be
satisfied at positive photon-exciton detunings in the so-called
thermodynamic condensation regime \cite{Kasprzak2008, Levrat2010}.

Condition (\ref{EqWlessG}) does not contradict the well-established
importance of interparticle interaction in polariton relaxation for
relatively high excitation densities for the noncondensed regime
\cite{Tartakovskii2000,Belykh2011,Bajoni2006, Qarry2003}. Indeed, in
the rate equation the income term describing polariton scattering
assisted by interparticle interaction is $w_\text{xeh}(1+\langle n
\rangle)$ whereas the corresponding term describing polariton escape
is $\gamma_\text{xeh} \langle n \rangle$. For $\langle n \rangle \ll
1$, the income term can dominate despite the condition
(\ref{EqWlessG}). The situation is reversed for the regime of
condensation, when $\langle n \rangle \gg 1$. Furthermore, our
conclusion does not contradict the reported enhancement of polariton
relaxation due to reservoir heating \cite{Tartakovskii2003} because
the value of $w_\text{xeh}$ grows with the reservoir temperature
[Eqs.~(\ref{EqWx}), (\ref{EqWe}), and~(\ref{EqDetailEq})].

\section{Conclusion}

We have studied theoretically and experimentally the polariton
linewidth and have shown that it is determined by the polariton
escape rate and polariton population. In experiments with resonant
excitation, the dynamics of the polariton linewidth is mainly
governed by the dynamics of the occupation number. By contrast, in
experiments with nonresonant excitation, this dynamics is mainly
governed by the dynamics of the polariton escape rate, which in turn
is governed by the dynamics of the reservoir temperature. On this
basis, we have developed a method of determining the reservoir
temperature by tracing the dependence of the polariton escape rate
on the polariton energy. The extracted reservoir temperature for
nonresonant pulsed excitation of a GaAs microcavity decays from
$\sim 100$~K at $50-100$~ps to the lattice temperature in a time of
$\sim 1$~ns. Increasing the excitation power leads to a slowdown of
the reservoir temperature relaxation. We have concluded that, in
experiments with nonresonant pulsed excitation of GaAs
microcavities, the reservoir temperature greatly exceeds the lattice
temperature in the regime of the polariton Bose-Einstein
condensation. As a result, the condensation is governed by the
phonon-assisted polariton relaxation, while the overall effect of
interparticle scattering is depopulation of the condensate.

\begin{acknowledgements}
We are grateful to N.~A.~Gippius, M.~V.~Kochiev, D.~A.~Mylnikov,
N.~N.~Sibeldin, M.~L.~Skorikov, and V.~A.~Tsvetkov for valuable
advice and useful discussions. The work is supported by the Russian
Foundation for Basic Research (Projects No.~12-02-33091,
No.~13-02-12197, No.~14-02-01073) and the Russian Academy of
Sciences. V.V.B. acknowledges support from the Russian Federation
President Scholarship.
\end{acknowledgements}

\providecommand{\noopsort}[1]{}\providecommand{\singleletter}[1]{#1}%


\begin{thebibliography}{52}%
\makeatletter
\providecommand \@ifxundefined [1]{%
 \@ifx{#1\undefined}
}%
\providecommand \@ifnum [1]{%
 \ifnum #1\expandafter \@firstoftwo
 \else \expandafter \@secondoftwo
 \fi
}%
\providecommand \@ifx [1]{%
 \ifx #1\expandafter \@firstoftwo
 \else \expandafter \@secondoftwo
 \fi
}%
\providecommand \natexlab [1]{#1}%
\providecommand \enquote  [1]{``#1''}%
\providecommand \bibnamefont  [1]{#1}%
\providecommand \bibfnamefont [1]{#1}%
\providecommand \citenamefont [1]{#1}%
\providecommand \href@noop [0]{\@secondoftwo}%
\providecommand \href [0]{\begingroup \@sanitize@url \@href}%
\providecommand \@href[1]{\@@startlink{#1}\@@href}%
\providecommand \@@href[1]{\endgroup#1\@@endlink}%
\providecommand \@sanitize@url [0]{\catcode `\\12\catcode
`\$12\catcode
  `\&12\catcode `\#12\catcode `\^12\catcode `\_12\catcode `\%12\relax}%
\providecommand \@@startlink[1]{}%
\providecommand \@@endlink[0]{}%
\providecommand \url  [0]{\begingroup\@sanitize@url \@url }%
\providecommand \@url [1]{\endgroup\@href {#1}{\urlprefix }}%
\providecommand \urlprefix  [0]{URL }%
\providecommand \Eprint [0]{\href }%
\providecommand \doibase [0]{http://dx.doi.org/}%
\providecommand \selectlanguage [0]{\@gobble}%
\providecommand \bibinfo  [0]{\@secondoftwo}%
\providecommand \bibfield  [0]{\@secondoftwo}%
\providecommand \translation [1]{[#1]}%
\providecommand \BibitemOpen [0]{}%
\providecommand \bibitemStop [0]{}%
\providecommand \bibitemNoStop [0]{.\EOS\space}%
\providecommand \EOS [0]{\spacefactor3000\relax}%
\providecommand \BibitemShut  [1]{\csname bibitem#1\endcsname}%
\let\auto@bib@innerbib\@empty
\bibitem [{\citenamefont {Shah}(1999)}]{Shah1999}%
  \BibitemOpen
  \bibfield  {author} {\bibinfo {author} {\bibfnamefont {J.}~\bibnamefont
  {Shah}},\ }\href@noop {} {\emph {\bibinfo {title} {{Ultrafast Spectroscopy of
  Semiconductors and Semiconductor Nanostructures}}}},\ \bibinfo {edition}
  {2nd}\ ed.,\ edited by\ \bibinfo {editor} {\bibfnamefont {M.}~\bibnamefont
  {Cardona}}, \bibinfo {editor} {\bibfnamefont {K.}~\bibnamefont {von
  Klitzing}}, \bibinfo {editor} {\bibfnamefont {R.}~\bibnamefont {Merlin}}, \
  and\ \bibinfo {editor} {\bibfnamefont {H.-J.}\ \bibnamefont {Queisser}}\
  (\bibinfo  {publisher} {Springer},\ \bibinfo {address} {Berlin},\ \bibinfo
  {year} {1999})\BibitemShut {NoStop}%
\bibitem [{\citenamefont {Knox}\ \emph {et~al.}(1986)\citenamefont {Knox},
  \citenamefont {Hirlimann}, \citenamefont {Miller}, \citenamefont {Shah},
  \citenamefont {Chemla},\ and\ \citenamefont {Shank}}]{Knox1986}%
  \BibitemOpen
  \bibfield  {author} {\bibinfo {author} {\bibfnamefont {W.~H.}\ \bibnamefont
  {Knox}}, \bibinfo {author} {\bibfnamefont {C.}~\bibnamefont {Hirlimann}},
  \bibinfo {author} {\bibfnamefont {D.~A.~B.}\ \bibnamefont {Miller}}, \bibinfo
  {author} {\bibfnamefont {J.}~\bibnamefont {Shah}}, \bibinfo {author}
  {\bibfnamefont {D.~S.}\ \bibnamefont {Chemla}}, \ and\ \bibinfo {author}
  {\bibfnamefont {C.~V.}\ \bibnamefont {Shank}},\ }\href {\doibase
  10.1103/PhysRevLett.56.1191} {\bibfield  {journal} {\bibinfo  {journal}
  {Phys. Rev. Lett.}\ }\textbf {\bibinfo {volume} {56}},\ \bibinfo {pages}
  {1191} (\bibinfo {year} {1986})}\BibitemShut {NoStop}%
\bibitem [{\citenamefont {Knox}\ \emph {et~al.}(1988)\citenamefont {Knox},
  \citenamefont {Chemla}, \citenamefont {Livescu}, \citenamefont {Cunningham},\
  and\ \citenamefont {Henry}}]{Knox1988}%
  \BibitemOpen
  \bibfield  {author} {\bibinfo {author} {\bibfnamefont {W.~H.}\ \bibnamefont
  {Knox}}, \bibinfo {author} {\bibfnamefont {D.~S.}\ \bibnamefont {Chemla}},
  \bibinfo {author} {\bibfnamefont {G.}~\bibnamefont {Livescu}}, \bibinfo
  {author} {\bibfnamefont {J.~E.}\ \bibnamefont {Cunningham}}, \ and\ \bibinfo
  {author} {\bibfnamefont {J.~E.}\ \bibnamefont {Henry}},\ }\href {\doibase
  10.1103/PhysRevLett.61.1290} {\bibfield  {journal} {\bibinfo  {journal}
  {Phys. Rev. Lett.}\ }\textbf {\bibinfo {volume} {61}},\ \bibinfo {pages}
  {1290} (\bibinfo {year} {1988})}\BibitemShut {NoStop}%
\bibitem [{\citenamefont {Leo}\ \emph {et~al.}(1988{\natexlab{a}})\citenamefont
  {Leo}, \citenamefont {R\"{u}hle},\ and\ \citenamefont {Ploog}}]{Leo1988}%
  \BibitemOpen
  \bibfield  {author} {\bibinfo {author} {\bibfnamefont {K.}~\bibnamefont
  {Leo}}, \bibinfo {author} {\bibfnamefont {W.~W.}\ \bibnamefont {R\"{u}hle}},
  \ and\ \bibinfo {author} {\bibfnamefont {K.}~\bibnamefont {Ploog}},\ }\href
  {\doibase 10.1103/PhysRevB.38.1947} {\bibfield  {journal} {\bibinfo
  {journal} {Phys. Rev. B}\ }\textbf {\bibinfo {volume} {38}},\ \bibinfo
  {pages} {1947} (\bibinfo {year} {1988}{\natexlab{a}})}\BibitemShut {NoStop}%
\bibitem [{\citenamefont {Leo}\ \emph {et~al.}(1988{\natexlab{b}})\citenamefont
  {Leo}, \citenamefont {R\"{u}hle}, \citenamefont {Queisser},\ and\
  \citenamefont {Ploog}}]{Leo1988a}%
  \BibitemOpen
  \bibfield  {author} {\bibinfo {author} {\bibfnamefont {K.}~\bibnamefont
  {Leo}}, \bibinfo {author} {\bibfnamefont {W.~W.}\ \bibnamefont {R\"{u}hle}},
  \bibinfo {author} {\bibfnamefont {H.~J.}\ \bibnamefont {Queisser}}, \ and\
  \bibinfo {author} {\bibfnamefont {K.}~\bibnamefont {Ploog}},\ }\href
  {\doibase 10.1103/PhysRevB.37.7121} {\bibfield  {journal} {\bibinfo
  {journal} {Phys. Rev. B}\ }\textbf {\bibinfo {volume} {37}},\ \bibinfo
  {pages} {7121} (\bibinfo {year} {1988}{\natexlab{b}})}\BibitemShut {NoStop}%
\bibitem [{\citenamefont {Yoon}\ \emph {et~al.}(1996)\citenamefont {Yoon},
  \citenamefont {Wake},\ and\ \citenamefont {Wolfe}}]{Yoon1996}%
  \BibitemOpen
  \bibfield  {author} {\bibinfo {author} {\bibfnamefont {H.~W.}\ \bibnamefont
  {Yoon}}, \bibinfo {author} {\bibfnamefont {D.~R.}\ \bibnamefont {Wake}}, \
  and\ \bibinfo {author} {\bibfnamefont {J.~P.}\ \bibnamefont {Wolfe}},\ }\href
  {\doibase 10.1103/PhysRevB.54.2763} {\bibfield  {journal} {\bibinfo
  {journal} {Phys. Rev. B}\ }\textbf {\bibinfo {volume} {54}},\ \bibinfo
  {pages} {2763} (\bibinfo {year} {1996})}\BibitemShut {NoStop}%
\bibitem [{\citenamefont {Rosenwaks}\ \emph {et~al.}(1993)\citenamefont
  {Rosenwaks}, \citenamefont {Hanna}, \citenamefont {Levi}, \citenamefont
  {Szmyd}, \citenamefont {Ahrenkiel},\ and\ \citenamefont
  {Nozik}}]{Rosenwaks1993}%
  \BibitemOpen
  \bibfield  {author} {\bibinfo {author} {\bibfnamefont {Y.}~\bibnamefont
  {Rosenwaks}}, \bibinfo {author} {\bibfnamefont {M.~C.}\ \bibnamefont
  {Hanna}}, \bibinfo {author} {\bibfnamefont {D.~H.}\ \bibnamefont {Levi}},
  \bibinfo {author} {\bibfnamefont {D.~M.}\ \bibnamefont {Szmyd}}, \bibinfo
  {author} {\bibfnamefont {R.~K.}\ \bibnamefont {Ahrenkiel}}, \ and\ \bibinfo
  {author} {\bibfnamefont {A.~J.}\ \bibnamefont {Nozik}},\ }\href {\doibase
  10.1103/PhysRevB.48.14675} {\bibfield  {journal} {\bibinfo  {journal} {Phys.
  Rev. B}\ }\textbf {\bibinfo {volume} {48}},\ \bibinfo {pages} {14675}
  (\bibinfo {year} {1993})}\BibitemShut {NoStop}%
\bibitem [{\citenamefont {Szczytko}\ \emph {et~al.}(2004)\citenamefont
  {Szczytko}, \citenamefont {Kappei}, \citenamefont {Berney}, \citenamefont
  {Morier-Genoud}, \citenamefont {Portella-Oberli},\ and\ \citenamefont
  {Deveaud}}]{Szczytko2004}%
  \BibitemOpen
  \bibfield  {author} {\bibinfo {author} {\bibfnamefont {J.}~\bibnamefont
  {Szczytko}}, \bibinfo {author} {\bibfnamefont {L.}~\bibnamefont {Kappei}},
  \bibinfo {author} {\bibfnamefont {J.}~\bibnamefont {Berney}}, \bibinfo
  {author} {\bibfnamefont {F.}~\bibnamefont {Morier-Genoud}}, \bibinfo {author}
  {\bibfnamefont {M.~T.}\ \bibnamefont {Portella-Oberli}}, \ and\ \bibinfo
  {author} {\bibfnamefont {B.}~\bibnamefont {Deveaud}},\ }\href {\doibase
  10.1103/PhysRevLett.93.137401} {\bibfield  {journal} {\bibinfo  {journal}
  {Phys. Rev. Lett.}\ }\textbf {\bibinfo {volume} {93}},\ \bibinfo {pages}
  {137401} (\bibinfo {year} {2004})}\BibitemShut {NoStop}%
\bibitem [{\citenamefont {Deveaud}\ \emph {et~al.}(1988)\citenamefont
  {Deveaud}, \citenamefont {Shah}, \citenamefont {Damen},\ and\ \citenamefont
  {Tsang}}]{Deveaud1988}%
  \BibitemOpen
  \bibfield  {author} {\bibinfo {author} {\bibfnamefont {B.}~\bibnamefont
  {Deveaud}}, \bibinfo {author} {\bibfnamefont {J.}~\bibnamefont {Shah}},
  \bibinfo {author} {\bibfnamefont {T.~C.}\ \bibnamefont {Damen}}, \ and\
  \bibinfo {author} {\bibfnamefont {W.~T.}\ \bibnamefont {Tsang}},\ }\href
  {\doibase 10.1063/1.99614} {\bibfield  {journal} {\bibinfo  {journal} {Appl.
  Phys. Lett.}\ }\textbf {\bibinfo {volume} {52}},\ \bibinfo {pages} {1886}
  (\bibinfo {year} {1988})}\BibitemShut {NoStop}%
\bibitem [{\citenamefont {Barros}\ \emph {et~al.}(1993)\citenamefont {Barros},
  \citenamefont {Becker}, \citenamefont {Morris}, \citenamefont {Deveaud},
  \citenamefont {Regreny},\ and\ \citenamefont {Beisser}}]{Barros1993}%
  \BibitemOpen
  \bibfield  {author} {\bibinfo {author} {\bibfnamefont {M.~R.~X.}\
  \bibnamefont {Barros}}, \bibinfo {author} {\bibfnamefont {P.~C.}\
  \bibnamefont {Becker}}, \bibinfo {author} {\bibfnamefont {D.}~\bibnamefont
  {Morris}}, \bibinfo {author} {\bibfnamefont {B.}~\bibnamefont {Deveaud}},
  \bibinfo {author} {\bibfnamefont {A.}~\bibnamefont {Regreny}}, \ and\
  \bibinfo {author} {\bibfnamefont {F.}~\bibnamefont {Beisser}},\ }\href
  {\doibase 10.1103/PhysRevB.47.10951} {\bibfield  {journal} {\bibinfo
  {journal} {Phys. Rev. B}\ }\textbf {\bibinfo {volume} {47}},\ \bibinfo
  {pages} {10951} (\bibinfo {year} {1993})}\BibitemShut {NoStop}%
\bibitem [{\citenamefont {Oberli}\ \emph {et~al.}(1989)\citenamefont {Oberli},
  \citenamefont {Shah}, \citenamefont {Jewell}, \citenamefont {Damen},\ and\
  \citenamefont {Chand}}]{Oberli1989}%
  \BibitemOpen
  \bibfield  {author} {\bibinfo {author} {\bibfnamefont {D.~Y.}\ \bibnamefont
  {Oberli}}, \bibinfo {author} {\bibfnamefont {J.}~\bibnamefont {Shah}},
  \bibinfo {author} {\bibfnamefont {J.~L.}\ \bibnamefont {Jewell}}, \bibinfo
  {author} {\bibfnamefont {T.~C.}\ \bibnamefont {Damen}}, \ and\ \bibinfo
  {author} {\bibfnamefont {N.}~\bibnamefont {Chand}},\ }\href {\doibase
  10.1063/1.100788} {\bibfield  {journal} {\bibinfo  {journal} {Appl. Phys.
  Lett.}\ }\textbf {\bibinfo {volume} {54}},\ \bibinfo {pages} {1028} (\bibinfo
  {year} {1989})}\BibitemShut {NoStop}%
\bibitem [{\citenamefont {Sotirelis}\ and\ \citenamefont
  {Hess}(1994)}]{Sotirelis1994}%
  \BibitemOpen
  \bibfield  {author} {\bibinfo {author} {\bibfnamefont {P.}~\bibnamefont
  {Sotirelis}}\ and\ \bibinfo {author} {\bibfnamefont {K.}~\bibnamefont
  {Hess}},\ }\href {\doibase 10.1103/PhysRevB.49.7543} {\bibfield  {journal}
  {\bibinfo  {journal} {Phys. Rev. B}\ }\textbf {\bibinfo {volume} {49}},\
  \bibinfo {pages} {7543} (\bibinfo {year} {1994})}\BibitemShut {NoStop}%
\bibitem [{\citenamefont {Hoyer}\ \emph {et~al.}(2005)\citenamefont {Hoyer},
  \citenamefont {Ell}, \citenamefont {Kira}, \citenamefont {Koch},
  \citenamefont {Chatterjee}, \citenamefont {Mosor}, \citenamefont {Khitrova},
  \citenamefont {Gibbs},\ and\ \citenamefont {Stolz}}]{Hoyer2005}%
  \BibitemOpen
  \bibfield  {author} {\bibinfo {author} {\bibfnamefont {W.}~\bibnamefont
  {Hoyer}}, \bibinfo {author} {\bibfnamefont {C.}~\bibnamefont {Ell}}, \bibinfo
  {author} {\bibfnamefont {M.}~\bibnamefont {Kira}}, \bibinfo {author}
  {\bibfnamefont {S.~W.}\ \bibnamefont {Koch}}, \bibinfo {author}
  {\bibfnamefont {S.}~\bibnamefont {Chatterjee}}, \bibinfo {author}
  {\bibfnamefont {S.}~\bibnamefont {Mosor}}, \bibinfo {author} {\bibfnamefont
  {G.}~\bibnamefont {Khitrova}}, \bibinfo {author} {\bibfnamefont {H.~M.}\
  \bibnamefont {Gibbs}}, \ and\ \bibinfo {author} {\bibfnamefont
  {H.}~\bibnamefont {Stolz}},\ }\href {\doibase 10.1103/PhysRevB.72.075324}
  {\bibfield  {journal} {\bibinfo  {journal} {Phys. Rev. B}\ }\textbf {\bibinfo
  {volume} {72}},\ \bibinfo {pages} {075324} (\bibinfo {year}
  {2005})}\BibitemShut {NoStop}%
\bibitem [{\citenamefont {Bagaev}\ \emph {et~al.}(2010)\citenamefont {Bagaev},
  \citenamefont {Krivobok}, \citenamefont {Nikolaev}, \citenamefont {Novikov},
  \citenamefont {Onishchenko},\ and\ \citenamefont {Skorikov}}]{Bagaev2010}%
  \BibitemOpen
  \bibfield  {author} {\bibinfo {author} {\bibfnamefont {V.~S.}\ \bibnamefont
  {Bagaev}}, \bibinfo {author} {\bibfnamefont {V.~S.}\ \bibnamefont
  {Krivobok}}, \bibinfo {author} {\bibfnamefont {S.~N.}\ \bibnamefont
  {Nikolaev}}, \bibinfo {author} {\bibfnamefont {A.~V.}\ \bibnamefont
  {Novikov}}, \bibinfo {author} {\bibfnamefont {E.~E.}\ \bibnamefont
  {Onishchenko}}, \ and\ \bibinfo {author} {\bibfnamefont {M.~L.}\ \bibnamefont
  {Skorikov}},\ }\href {\doibase 10.1103/PhysRevB.82.115313} {\bibfield
  {journal} {\bibinfo  {journal} {Phys. Rev. B}\ }\textbf {\bibinfo {volume}
  {82}},\ \bibinfo {pages} {115313} (\bibinfo {year} {2010})}\BibitemShut
  {NoStop}%
\bibitem [{\citenamefont {Kasprzak}\ \emph {et~al.}(2006)\citenamefont
  {Kasprzak}, \citenamefont {Richard}, \citenamefont {Kundermann},
  \citenamefont {Baas}, \citenamefont {Jeambrun}, \citenamefont {Keeling},
  \citenamefont {Marchetti}, \citenamefont {Szymanska}, \citenamefont
  {Andr\'{e}}, \citenamefont {Staehli}, \citenamefont {Savona}, \citenamefont
  {Littlewood}, \citenamefont {Deveaud},\ and\ \citenamefont
  {Dang}}]{Kasprzak2006}%
  \BibitemOpen
  \bibfield  {author} {\bibinfo {author} {\bibfnamefont {J.}~\bibnamefont
  {Kasprzak}}, \bibinfo {author} {\bibfnamefont {M.}~\bibnamefont {Richard}},
  \bibinfo {author} {\bibfnamefont {S.}~\bibnamefont {Kundermann}}, \bibinfo
  {author} {\bibfnamefont {A.}~\bibnamefont {Baas}}, \bibinfo {author}
  {\bibfnamefont {P.}~\bibnamefont {Jeambrun}}, \bibinfo {author}
  {\bibfnamefont {J.~M.~J.}\ \bibnamefont {Keeling}}, \bibinfo {author}
  {\bibfnamefont {F.~M.}\ \bibnamefont {Marchetti}}, \bibinfo {author}
  {\bibfnamefont {M.~H.}\ \bibnamefont {Szymanska}}, \bibinfo {author}
  {\bibfnamefont {R.}~\bibnamefont {Andr\'{e}}}, \bibinfo {author}
  {\bibfnamefont {J.~L.}\ \bibnamefont {Staehli}}, \bibinfo {author}
  {\bibfnamefont {V.}~\bibnamefont {Savona}}, \bibinfo {author} {\bibfnamefont
  {P.~B.}\ \bibnamefont {Littlewood}}, \bibinfo {author} {\bibfnamefont
  {B.}~\bibnamefont {Deveaud}}, \ and\ \bibinfo {author} {\bibfnamefont
  {L.~S.}\ \bibnamefont {Dang}},\ }\href {\doibase 10.1038/nature05131}
  {\bibfield  {journal} {\bibinfo  {journal} {Nature (London)}\ }\textbf
  {\bibinfo {volume} {443}},\ \bibinfo {pages} {409} (\bibinfo {year}
  {2006})}\BibitemShut {NoStop}%
\bibitem [{\citenamefont {Sanvitto}\ and\ \citenamefont
  {Timofeev}(2012)}]{Sanvitto2012}%
  \BibitemOpen
  \bibfield  {author} {\bibinfo {author} {\bibfnamefont {D.}~\bibnamefont
  {Sanvitto}}\ and\ \bibinfo {author} {\bibfnamefont {V.}~\bibnamefont
  {Timofeev}},\ }\href {\doibase 10.1007/978-3-642-24186-4} {\emph {\bibinfo
  {title} {{Exciton Polaritons in Microcavities}}}},\ edited by\ \bibinfo
  {editor} {\bibfnamefont {V.}~\bibnamefont {Timofeev}}\ and\ \bibinfo {editor}
  {\bibfnamefont {D.}~\bibnamefont {Sanvitto}},\ \bibinfo {series} {Springer
  Series in Solid-State Sciences}, Vol.\ \bibinfo {volume} {172}\ (\bibinfo
  {publisher} {Springer},\ \bibinfo {address} {Berlin},\ \bibinfo {year}
  {2012})\BibitemShut {NoStop}%
\bibitem [{\citenamefont {Deng}\ \emph {et~al.}(2003)\citenamefont {Deng},
  \citenamefont {Weihs}, \citenamefont {Snoke}, \citenamefont {Bloch},\ and\
  \citenamefont {Yamamoto}}]{Deng2003}%
  \BibitemOpen
  \bibfield  {author} {\bibinfo {author} {\bibfnamefont {H.}~\bibnamefont
  {Deng}}, \bibinfo {author} {\bibfnamefont {G.}~\bibnamefont {Weihs}},
  \bibinfo {author} {\bibfnamefont {D.}~\bibnamefont {Snoke}}, \bibinfo
  {author} {\bibfnamefont {J.}~\bibnamefont {Bloch}}, \ and\ \bibinfo {author}
  {\bibfnamefont {Y.}~\bibnamefont {Yamamoto}},\ }\href {\doibase
  10.1073/pnas.2634328100} {\bibfield  {journal} {\bibinfo  {journal} {Proc.
  Natl. Acad. Sci. U. S. A.}\ }\textbf {\bibinfo {volume} {100}},\ \bibinfo
  {pages} {15318} (\bibinfo {year} {2003})}\BibitemShut {NoStop}%
\bibitem [{\citenamefont {Deng}\ \emph {et~al.}(2006)\citenamefont {Deng},
  \citenamefont {Press}, \citenamefont {G\"{o}tzinger}, \citenamefont
  {Solomon}, \citenamefont {Hey}, \citenamefont {Ploog},\ and\ \citenamefont
  {Yamamoto}}]{Deng2006}%
  \BibitemOpen
  \bibfield  {author} {\bibinfo {author} {\bibfnamefont {H.}~\bibnamefont
  {Deng}}, \bibinfo {author} {\bibfnamefont {D.}~\bibnamefont {Press}},
  \bibinfo {author} {\bibfnamefont {S.}~\bibnamefont {G\"{o}tzinger}}, \bibinfo
  {author} {\bibfnamefont {G.}~\bibnamefont {Solomon}}, \bibinfo {author}
  {\bibfnamefont {R.}~\bibnamefont {Hey}}, \bibinfo {author} {\bibfnamefont
  {K.~H.}\ \bibnamefont {Ploog}}, \ and\ \bibinfo {author} {\bibfnamefont
  {Y.}~\bibnamefont {Yamamoto}},\ }\href {\doibase
  10.1103/PhysRevLett.97.146402} {\bibfield  {journal} {\bibinfo  {journal}
  {Phys. Rev. Lett.}\ }\textbf {\bibinfo {volume} {97}},\ \bibinfo {pages}
  {146402} (\bibinfo {year} {2006})}\BibitemShut {NoStop}%
\bibitem [{\citenamefont {Balili}\ \emph {et~al.}(2007)\citenamefont {Balili},
  \citenamefont {Hartwell}, \citenamefont {Snoke}, \citenamefont {Pfeiffer},\
  and\ \citenamefont {West}}]{Balili2007}%
  \BibitemOpen
  \bibfield  {author} {\bibinfo {author} {\bibfnamefont {R.}~\bibnamefont
  {Balili}}, \bibinfo {author} {\bibfnamefont {V.}~\bibnamefont {Hartwell}},
  \bibinfo {author} {\bibfnamefont {D.}~\bibnamefont {Snoke}}, \bibinfo
  {author} {\bibfnamefont {L.}~\bibnamefont {Pfeiffer}}, \ and\ \bibinfo
  {author} {\bibfnamefont {K.}~\bibnamefont {West}},\ }\href {\doibase
  10.1126/science.1140990} {\bibfield  {journal} {\bibinfo  {journal}
  {Science}\ }\textbf {\bibinfo {volume} {316}},\ \bibinfo {pages} {1007}
  (\bibinfo {year} {2007})}\BibitemShut {NoStop}%
\bibitem [{\citenamefont {Kasprzak}\ \emph {et~al.}(2008)\citenamefont
  {Kasprzak}, \citenamefont {Solnyshkov}, \citenamefont {Andr\'{e}},
  \citenamefont {Dang},\ and\ \citenamefont {Malpuech}}]{Kasprzak2008}%
  \BibitemOpen
  \bibfield  {author} {\bibinfo {author} {\bibfnamefont {J.}~\bibnamefont
  {Kasprzak}}, \bibinfo {author} {\bibfnamefont {D.~D.}\ \bibnamefont
  {Solnyshkov}}, \bibinfo {author} {\bibfnamefont {R.}~\bibnamefont
  {Andr\'{e}}}, \bibinfo {author} {\bibfnamefont {L.~S.}\ \bibnamefont {Dang}},
  \ and\ \bibinfo {author} {\bibfnamefont {G.}~\bibnamefont {Malpuech}},\
  }\href {\doibase 10.1103/PhysRevLett.101.146404} {\bibfield  {journal}
  {\bibinfo  {journal} {Phys. Rev. Lett.}\ }\textbf {\bibinfo {volume} {101}},\
  \bibinfo {pages} {146404} (\bibinfo {year} {2008})}\BibitemShut {NoStop}%
\bibitem [{\citenamefont {Levrat}\ \emph {et~al.}(2010)\citenamefont {Levrat},
  \citenamefont {Butt\'{e}}, \citenamefont {Feltin}, \citenamefont {Carlin},
  \citenamefont {Grandjean}, \citenamefont {Solnyshkov},\ and\ \citenamefont
  {Malpuech}}]{Levrat2010}%
  \BibitemOpen
  \bibfield  {author} {\bibinfo {author} {\bibfnamefont {J.}~\bibnamefont
  {Levrat}}, \bibinfo {author} {\bibfnamefont {R.}~\bibnamefont {Butt\'{e}}},
  \bibinfo {author} {\bibfnamefont {E.}~\bibnamefont {Feltin}}, \bibinfo
  {author} {\bibfnamefont {J.-F.}\ \bibnamefont {Carlin}}, \bibinfo {author}
  {\bibfnamefont {N.}~\bibnamefont {Grandjean}}, \bibinfo {author}
  {\bibfnamefont {D.}~\bibnamefont {Solnyshkov}}, \ and\ \bibinfo {author}
  {\bibfnamefont {G.}~\bibnamefont {Malpuech}},\ }\href {\doibase
  10.1103/PhysRevB.81.125305} {\bibfield  {journal} {\bibinfo  {journal} {Phys.
  Rev. B}\ }\textbf {\bibinfo {volume} {81}},\ \bibinfo {pages} {125305}
  (\bibinfo {year} {2010})}\BibitemShut {NoStop}%
\bibitem [{\citenamefont {Kammann}\ \emph {et~al.}(2012)\citenamefont
  {Kammann}, \citenamefont {Ohadi}, \citenamefont {Maragkou}, \citenamefont
  {Kavokin},\ and\ \citenamefont {Lagoudakis}}]{Kammann2012}%
  \BibitemOpen
  \bibfield  {author} {\bibinfo {author} {\bibfnamefont {E.}~\bibnamefont
  {Kammann}}, \bibinfo {author} {\bibfnamefont {H.}~\bibnamefont {Ohadi}},
  \bibinfo {author} {\bibfnamefont {M.}~\bibnamefont {Maragkou}}, \bibinfo
  {author} {\bibfnamefont {A.~V.}\ \bibnamefont {Kavokin}}, \ and\ \bibinfo
  {author} {\bibfnamefont {P.~G.}\ \bibnamefont {Lagoudakis}},\ }\href
  {\doibase 10.1088/1367-2630/14/10/105003} {\bibfield  {journal} {\bibinfo
  {journal} {New J. Phys.}\ }\textbf {\bibinfo {volume} {14}},\ \bibinfo
  {pages} {105003} (\bibinfo {year} {2012})}\BibitemShut {NoStop}%
\bibitem [{\citenamefont {Bajoni}\ \emph {et~al.}(2006)\citenamefont {Bajoni},
  \citenamefont {Perrin}, \citenamefont {Senellart}, \citenamefont
  {Lema\^{\i}tre}, \citenamefont {Sermage},\ and\ \citenamefont
  {Bloch}}]{Bajoni2006}%
  \BibitemOpen
  \bibfield  {author} {\bibinfo {author} {\bibfnamefont {D.}~\bibnamefont
  {Bajoni}}, \bibinfo {author} {\bibfnamefont {M.}~\bibnamefont {Perrin}},
  \bibinfo {author} {\bibfnamefont {P.}~\bibnamefont {Senellart}}, \bibinfo
  {author} {\bibfnamefont {A.}~\bibnamefont {Lema\^{\i}tre}}, \bibinfo {author}
  {\bibfnamefont {B.}~\bibnamefont {Sermage}}, \ and\ \bibinfo {author}
  {\bibfnamefont {J.}~\bibnamefont {Bloch}},\ }\href {\doibase
  10.1103/PhysRevB.73.205344} {\bibfield  {journal} {\bibinfo  {journal} {Phys.
  Rev. B}\ }\textbf {\bibinfo {volume} {73}},\ \bibinfo {pages} {205344}
  (\bibinfo {year} {2006})}\BibitemShut {NoStop}%
\bibitem [{\citenamefont {Loudon}(2000)}]{Loudon2000}%
  \BibitemOpen
  \bibfield  {author} {\bibinfo {author} {\bibfnamefont {R.}~\bibnamefont
  {Loudon}},\ }\href@noop {} {\emph {\bibinfo {title} {The Quantum Theory of
  Light}}},\ \bibinfo {edition} {3rd}\ ed.\ (\bibinfo  {publisher} {Oxford
  University Press},\ \bibinfo {address} {New York},\ \bibinfo {year}
  {2000})\BibitemShut {NoStop}%
\bibitem [{\citenamefont {Wiener}(1930)}]{Wiener1930}%
  \BibitemOpen
  \bibfield  {author} {\bibinfo {author} {\bibfnamefont {N.}~\bibnamefont
  {Wiener}},\ }\href@noop {} {\bibfield  {journal} {\bibinfo  {journal} {Acta
  Math.}\ }\textbf {\bibinfo {volume} {55}},\ \bibinfo {pages} {117} (\bibinfo
  {year} {1930})}\BibitemShut {NoStop}%
\bibitem [{\citenamefont {Khintchine}(1934)}]{Khintchine1934}%
  \BibitemOpen
  \bibfield  {author} {\bibinfo {author} {\bibfnamefont {A.}~\bibnamefont
  {Khintchine}},\ }\href@noop {} {\bibfield  {journal} {\bibinfo  {journal}
  {Math. Ann.}\ }\textbf {\bibinfo {volume} {109}},\ \bibinfo {pages} {604}
  (\bibinfo {year} {1934})}\BibitemShut {NoStop}%
\bibitem [{\citenamefont {Einstein}(1914)}]{Einstein1914}%
  \BibitemOpen
  \bibfield  {author} {\bibinfo {author} {\bibfnamefont {A.}~\bibnamefont
  {Einstein}},\ }\href@noop {} {\bibfield  {journal} {\bibinfo  {journal}
  {Arch. Sci. Phys. Nat.}\ }\textbf {\bibinfo {volume} {37}},\ \bibinfo {pages}
  {254} (\bibinfo {year} {1914})}\BibitemShut {NoStop}%
\bibitem [{\citenamefont {Yaglom}(1987)}]{Yaglom1987}%
  \BibitemOpen
  \bibfield  {author} {\bibinfo {author} {\bibfnamefont {A.~M.}\ \bibnamefont
  {Yaglom}},\ }\href@noop {} {\bibfield  {journal} {\bibinfo  {journal} {IEEE
  ASSP Mag.}\ }\textbf {\bibinfo {volume} {4}},\ \bibinfo {pages} {7} (\bibinfo
  {year} {1987})}\BibitemShut {NoStop}%
\bibitem [{\citenamefont {Lax}(1963)}]{Lax1963}%
  \BibitemOpen
  \bibfield  {author} {\bibinfo {author} {\bibfnamefont {M.}~\bibnamefont
  {Lax}},\ }\href@noop {} {\bibfield  {journal} {\bibinfo  {journal} {Phys.
  Rev.}\ }\textbf {\bibinfo {volume} {129}},\ \bibinfo {pages} {2342} (\bibinfo
  {year} {1963})}\BibitemShut {NoStop}%
\bibitem [{\citenamefont {Lax}(1967)}]{Lax1967}%
  \BibitemOpen
  \bibfield  {author} {\bibinfo {author} {\bibfnamefont {M.}~\bibnamefont
  {Lax}},\ }\href@noop {} {\bibfield  {journal} {\bibinfo  {journal} {Phys.
  Rev.}\ }\textbf {\bibinfo {volume} {157}},\ \bibinfo {pages} {213} (\bibinfo
  {year} {1967})}\BibitemShut {NoStop}%
\bibitem [{\citenamefont {Carmichael}(1993)}]{Carmichael1993}%
  \BibitemOpen
  \bibfield  {author} {\bibinfo {author} {\bibfnamefont {H.}~\bibnamefont
  {Carmichael}},\ }\href@noop {} {\emph {\bibinfo {title} {An Open Systems
  Approach to Quantum Optics}}}\ (\bibinfo  {publisher} {Springer},\ \bibinfo
  {address} {Berlin},\ \bibinfo {year} {1993})\BibitemShut {NoStop}%
\bibitem [{\citenamefont {Porras}\ and\ \citenamefont
  {Tejedor}(2003)}]{PorrasTejedor2003}%
  \BibitemOpen
  \bibfield  {author} {\bibinfo {author} {\bibfnamefont {D.}~\bibnamefont
  {Porras}}\ and\ \bibinfo {author} {\bibfnamefont {C.}~\bibnamefont
  {Tejedor}},\ }\href@noop {} {\bibfield  {journal} {\bibinfo  {journal} {Phys.
  Rev. B}\ }\textbf {\bibinfo {volume} {67}},\ \bibinfo {pages} {161310(R)}
  (\bibinfo {year} {2003})}\BibitemShut {NoStop}%
\bibitem [{\citenamefont {Rota}\ \emph {et~al.}(1993)\citenamefont {Rota},
  \citenamefont {Lugli}, \citenamefont {Elsaesser},\ and\ \citenamefont
  {Shah}}]{Rota1993}%
  \BibitemOpen
  \bibfield  {author} {\bibinfo {author} {\bibfnamefont {L.}~\bibnamefont
  {Rota}}, \bibinfo {author} {\bibfnamefont {P.}~\bibnamefont {Lugli}},
  \bibinfo {author} {\bibfnamefont {T.}~\bibnamefont {Elsaesser}}, \ and\
  \bibinfo {author} {\bibfnamefont {J.}~\bibnamefont {Shah}},\ }\href {\doibase
  10.1103/PhysRevB.47.4226} {\bibfield  {journal} {\bibinfo  {journal} {Phys.
  Rev. B}\ }\textbf {\bibinfo {volume} {47}},\ \bibinfo {pages} {4226}
  (\bibinfo {year} {1993})}\BibitemShut {NoStop}%
\bibitem [{\citenamefont {Alexandrou}\ \emph {et~al.}(1995)\citenamefont
  {Alexandrou}, \citenamefont {Berger},\ and\ \citenamefont
  {Hulin}}]{Alexandrou1995}%
  \BibitemOpen
  \bibfield  {author} {\bibinfo {author} {\bibfnamefont {A.}~\bibnamefont
  {Alexandrou}}, \bibinfo {author} {\bibfnamefont {V.}~\bibnamefont {Berger}},
  \ and\ \bibinfo {author} {\bibfnamefont {D.}~\bibnamefont {Hulin}},\ }\href
  {\doibase 10.1103/PhysRevB.52.4654} {\bibfield  {journal} {\bibinfo
  {journal} {Phys. Rev. B}\ }\textbf {\bibinfo {volume} {52}},\ \bibinfo
  {pages} {4654} (\bibinfo {year} {1995})}\BibitemShut {NoStop}%
\bibitem [{\citenamefont {Tassone}\ and\ \citenamefont
  {Yamamoto}(1999)}]{Tassone1999}%
  \BibitemOpen
  \bibfield  {author} {\bibinfo {author} {\bibfnamefont {F.}~\bibnamefont
  {Tassone}}\ and\ \bibinfo {author} {\bibfnamefont {Y.}~\bibnamefont
  {Yamamoto}},\ }\href {\doibase 10.1103/PhysRevB.59.10830} {\bibfield
  {journal} {\bibinfo  {journal} {Phys. Rev. B}\ }\textbf {\bibinfo {volume}
  {59}},\ \bibinfo {pages} {10830} (\bibinfo {year} {1999})}\BibitemShut
  {NoStop}%
\bibitem [{\citenamefont {Porras}\ \emph {et~al.}(2002)\citenamefont {Porras},
  \citenamefont {Ciuti}, \citenamefont {Baumberg},\ and\ \citenamefont
  {Tejedor}}]{Porras2002}%
  \BibitemOpen
  \bibfield  {author} {\bibinfo {author} {\bibfnamefont {D.}~\bibnamefont
  {Porras}}, \bibinfo {author} {\bibfnamefont {C.}~\bibnamefont {Ciuti}},
  \bibinfo {author} {\bibfnamefont {J.~J.}\ \bibnamefont {Baumberg}}, \ and\
  \bibinfo {author} {\bibfnamefont {C.}~\bibnamefont {Tejedor}},\ }\href
  {\doibase 10.1103/PhysRevB.66.085304} {\bibfield  {journal} {\bibinfo
  {journal} {Phys. Rev. B}\ }\textbf {\bibinfo {volume} {66}},\ \bibinfo
  {pages} {085304} (\bibinfo {year} {2002})}\BibitemShut {NoStop}%
\bibitem [{\citenamefont {Belykh}\ \emph {et~al.}(2011)\citenamefont {Belykh},
  \citenamefont {Tsvetkov}, \citenamefont {Skorikov},\ and\ \citenamefont
  {Sibeldin}}]{Belykh2011}%
  \BibitemOpen
  \bibfield  {author} {\bibinfo {author} {\bibfnamefont {V.~V.}\ \bibnamefont
  {Belykh}}, \bibinfo {author} {\bibfnamefont {V.~A.}\ \bibnamefont
  {Tsvetkov}}, \bibinfo {author} {\bibfnamefont {M.~L.}\ \bibnamefont
  {Skorikov}}, \ and\ \bibinfo {author} {\bibfnamefont {N.~N.}\ \bibnamefont
  {Sibeldin}},\ }\href@noop {} {\bibfield  {journal} {\bibinfo  {journal} {J.
  Phys. Condens. Matter}\ }\textbf {\bibinfo {volume} {23}},\ \bibinfo {pages}
  {215302} (\bibinfo {year} {2011})}\BibitemShut {NoStop}%
\bibitem [{\citenamefont {Belykh}\ \emph {et~al.}(2012)\citenamefont {Belykh},
  \citenamefont {Mylnikov},\ and\ \citenamefont {Sibeldin}}]{Belykh2012}%
  \BibitemOpen
  \bibfield  {author} {\bibinfo {author} {\bibfnamefont {V.~V.}\ \bibnamefont
  {Belykh}}, \bibinfo {author} {\bibfnamefont {D.~A.}\ \bibnamefont
  {Mylnikov}}, \ and\ \bibinfo {author} {\bibfnamefont {N.~N.}\ \bibnamefont
  {Sibeldin}},\ }\href {\doibase 10.1002/pssc.201100199} {\bibfield  {journal}
  {\bibinfo  {journal} {Phys. Status Solidi C}\ }\textbf {\bibinfo {volume}
  {9}},\ \bibinfo {pages} {1230} (\bibinfo {year} {2012})}\BibitemShut
  {NoStop}%
\bibitem [{\citenamefont {Renucci}\ \emph {et~al.}(2005)\citenamefont
  {Renucci}, \citenamefont {Amand}, \citenamefont {Marie}, \citenamefont
  {Senellart}, \citenamefont {Bloch}, \citenamefont {Sermage},\ and\
  \citenamefont {Kavokin}}]{Renucci2005}%
  \BibitemOpen
  \bibfield  {author} {\bibinfo {author} {\bibfnamefont {P.}~\bibnamefont
  {Renucci}}, \bibinfo {author} {\bibfnamefont {T.}~\bibnamefont {Amand}},
  \bibinfo {author} {\bibfnamefont {X.}~\bibnamefont {Marie}}, \bibinfo
  {author} {\bibfnamefont {P.}~\bibnamefont {Senellart}}, \bibinfo {author}
  {\bibfnamefont {J.}~\bibnamefont {Bloch}}, \bibinfo {author} {\bibfnamefont
  {B.}~\bibnamefont {Sermage}}, \ and\ \bibinfo {author} {\bibfnamefont
  {K.~V.}\ \bibnamefont {Kavokin}},\ }\href {\doibase
  10.1103/PhysRevB.72.075317} {\bibfield  {journal} {\bibinfo  {journal} {Phys.
  Rev. B}\ }\textbf {\bibinfo {volume} {72}},\ \bibinfo {pages} {075317}
  (\bibinfo {year} {2005})}\BibitemShut {NoStop}%
\bibitem [{\citenamefont {Tempel}\ \emph {et~al.}(2012)\citenamefont {Tempel},
  \citenamefont {Veit}, \citenamefont {A\ss~mann}, \citenamefont {Kreilkamp},
  \citenamefont {Rahimi-Iman}, \citenamefont {L\"{o}ffler}, \citenamefont
  {H\"{o}fling}, \citenamefont {Reitzenstein}, \citenamefont {Worschech},
  \citenamefont {Forchel},\ and\ \citenamefont {Bayer}}]{Tempel2012}%
  \BibitemOpen
  \bibfield  {author} {\bibinfo {author} {\bibfnamefont {J.-S.}\ \bibnamefont
  {Tempel}}, \bibinfo {author} {\bibfnamefont {F.}~\bibnamefont {Veit}},
  \bibinfo {author} {\bibfnamefont {M.}~\bibnamefont {A\ss~mann}}, \bibinfo
  {author} {\bibfnamefont {L.~E.}\ \bibnamefont {Kreilkamp}}, \bibinfo {author}
  {\bibfnamefont {A.}~\bibnamefont {Rahimi-Iman}}, \bibinfo {author}
  {\bibfnamefont {A.}~\bibnamefont {L\"{o}ffler}}, \bibinfo {author}
  {\bibfnamefont {S.}~\bibnamefont {H\"{o}fling}}, \bibinfo {author}
  {\bibfnamefont {S.}~\bibnamefont {Reitzenstein}}, \bibinfo {author}
  {\bibfnamefont {L.}~\bibnamefont {Worschech}}, \bibinfo {author}
  {\bibfnamefont {A.}~\bibnamefont {Forchel}}, \ and\ \bibinfo {author}
  {\bibfnamefont {M.}~\bibnamefont {Bayer}},\ }\href {\doibase
  10.1103/PhysRevB.85.075318} {\bibfield  {journal} {\bibinfo  {journal} {Phys.
  Rev. B}\ }\textbf {\bibinfo {volume} {85}},\ \bibinfo {pages} {075318}
  (\bibinfo {year} {2012})}\BibitemShut {NoStop}%
\bibitem [{\citenamefont {Belykh}\ \emph {et~al.}(2013)\citenamefont {Belykh},
  \citenamefont {Sibeldin}, \citenamefont {Kulakovskii}, \citenamefont
  {Glazov}, \citenamefont {Semina}, \citenamefont {Schneider}, \citenamefont
  {H\"{o}fling}, \citenamefont {Kamp},\ and\ \citenamefont
  {Forchel}}]{Belykh2013}%
  \BibitemOpen
  \bibfield  {author} {\bibinfo {author} {\bibfnamefont {V.~V.}\ \bibnamefont
  {Belykh}}, \bibinfo {author} {\bibfnamefont {N.~N.}\ \bibnamefont
  {Sibeldin}}, \bibinfo {author} {\bibfnamefont {V.~D.}\ \bibnamefont
  {Kulakovskii}}, \bibinfo {author} {\bibfnamefont {M.~M.}\ \bibnamefont
  {Glazov}}, \bibinfo {author} {\bibfnamefont {M.~A.}\ \bibnamefont {Semina}},
  \bibinfo {author} {\bibfnamefont {C.}~\bibnamefont {Schneider}}, \bibinfo
  {author} {\bibfnamefont {S.}~\bibnamefont {H\"{o}fling}}, \bibinfo {author}
  {\bibfnamefont {M.}~\bibnamefont {Kamp}}, \ and\ \bibinfo {author}
  {\bibfnamefont {A.}~\bibnamefont {Forchel}},\ }\href {\doibase
  10.1103/PhysRevLett.110.137402} {\bibfield  {journal} {\bibinfo  {journal}
  {Phys. Rev. Lett.}\ }\textbf {\bibinfo {volume} {110}},\ \bibinfo {pages}
  {137402} (\bibinfo {year} {2013})},\ \Eprint {http://arxiv.org/abs/1210.6906}
  {arXiv:1210.6906} \BibitemShut {NoStop}%
\bibitem [{\citenamefont {Houdr\'{e}}\ \emph {et~al.}(1994)\citenamefont
  {Houdr\'{e}}, \citenamefont {Stanley}, \citenamefont {Oesterle},
  \citenamefont {Ilegems},\ and\ \citenamefont {Weisbuch}}]{Houdre1994}%
  \BibitemOpen
  \bibfield  {author} {\bibinfo {author} {\bibfnamefont {R.}~\bibnamefont
  {Houdr\'{e}}}, \bibinfo {author} {\bibfnamefont {R.~P.}\ \bibnamefont
  {Stanley}}, \bibinfo {author} {\bibfnamefont {U.}~\bibnamefont {Oesterle}},
  \bibinfo {author} {\bibfnamefont {M.}~\bibnamefont {Ilegems}}, \ and\
  \bibinfo {author} {\bibfnamefont {C.}~\bibnamefont {Weisbuch}},\ }\href
  {\doibase 10.1103/PhysRevB.49.16761} {\bibfield  {journal} {\bibinfo
  {journal} {Phys. Rev. B}\ }\textbf {\bibinfo {volume} {49}},\ \bibinfo
  {pages} {16761} (\bibinfo {year} {1994})}\BibitemShut {NoStop}%
\bibitem [{\citenamefont {Tsintzos}\ \emph {et~al.}(2008)\citenamefont
  {Tsintzos}, \citenamefont {Pelekanos}, \citenamefont {Konstantinidis},
  \citenamefont {Hatzopoulos},\ and\ \citenamefont {Savvidis}}]{Tsintzos2008}%
  \BibitemOpen
  \bibfield  {author} {\bibinfo {author} {\bibfnamefont {S.~I.}\ \bibnamefont
  {Tsintzos}}, \bibinfo {author} {\bibfnamefont {N.~T.}\ \bibnamefont
  {Pelekanos}}, \bibinfo {author} {\bibfnamefont {G.}~\bibnamefont
  {Konstantinidis}}, \bibinfo {author} {\bibfnamefont {Z.}~\bibnamefont
  {Hatzopoulos}}, \ and\ \bibinfo {author} {\bibfnamefont {P.~G.}\ \bibnamefont
  {Savvidis}},\ }\href {\doibase 10.1038/nature06979} {\bibfield  {journal}
  {\bibinfo  {journal} {Nature (London)}\ }\textbf {\bibinfo {volume} {453}},\
  \bibinfo {pages} {372} (\bibinfo {year} {2008})}\BibitemShut {NoStop}%
\bibitem [{\citenamefont {Krizhanovskii}\ \emph {et~al.}(2007)\citenamefont
  {Krizhanovskii}, \citenamefont {Love}, \citenamefont {Sanvitto},
  \citenamefont {Whittaker}, \citenamefont {Skolnick},\ and\ \citenamefont
  {Roberts}}]{Krizhanovskii2007}%
  \BibitemOpen
  \bibfield  {author} {\bibinfo {author} {\bibfnamefont {D.~N.}\ \bibnamefont
  {Krizhanovskii}}, \bibinfo {author} {\bibfnamefont {A.~P.~D.}\ \bibnamefont
  {Love}}, \bibinfo {author} {\bibfnamefont {D.}~\bibnamefont {Sanvitto}},
  \bibinfo {author} {\bibfnamefont {D.~M.}\ \bibnamefont {Whittaker}}, \bibinfo
  {author} {\bibfnamefont {M.~S.}\ \bibnamefont {Skolnick}}, \ and\ \bibinfo
  {author} {\bibfnamefont {J.~S.}\ \bibnamefont {Roberts}},\ }\href {\doibase
  10.1103/PhysRevB.75.233307} {\bibfield  {journal} {\bibinfo  {journal} {Phys.
  Rev. B}\ }\textbf {\bibinfo {volume} {75}},\ \bibinfo {pages} {233307}
  (\bibinfo {year} {2007})}\BibitemShut {NoStop}%
\bibitem [{\citenamefont {Hillmer}\ \emph {et~al.}(1989)\citenamefont
  {Hillmer}, \citenamefont {Forchel}, \citenamefont {Hansmann}, \citenamefont
  {Morohashi}, \citenamefont {Lopez}, \citenamefont {Meier},\ and\
  \citenamefont {Ploog}}]{Hillmer1989}%
  \BibitemOpen
  \bibfield  {author} {\bibinfo {author} {\bibfnamefont {H.}~\bibnamefont
  {Hillmer}}, \bibinfo {author} {\bibfnamefont {A.}~\bibnamefont {Forchel}},
  \bibinfo {author} {\bibfnamefont {S.}~\bibnamefont {Hansmann}}, \bibinfo
  {author} {\bibfnamefont {M.}~\bibnamefont {Morohashi}}, \bibinfo {author}
  {\bibfnamefont {E.}~\bibnamefont {Lopez}}, \bibinfo {author} {\bibfnamefont
  {H.~P.}\ \bibnamefont {Meier}}, \ and\ \bibinfo {author} {\bibfnamefont
  {K.}~\bibnamefont {Ploog}},\ }\href {\doibase 10.1103/PhysRevB.39.10901}
  {\bibfield  {journal} {\bibinfo  {journal} {Phys. Rev. B}\ }\textbf {\bibinfo
  {volume} {39}},\ \bibinfo {pages} {10901} (\bibinfo {year}
  {1989})}\BibitemShut {NoStop}%
\bibitem [{\citenamefont {Wertz}\ \emph {et~al.}(2009)\citenamefont {Wertz},
  \citenamefont {Ferrier}, \citenamefont {Solnyshkov}, \citenamefont
  {Senellart}, \citenamefont {Bajoni}, \citenamefont {Miard}, \citenamefont
  {Lemaitre}, \citenamefont {Malpuech},\ and\ \citenamefont
  {Bloch}}]{Wertz2009}%
  \BibitemOpen
  \bibfield  {author} {\bibinfo {author} {\bibfnamefont {E.}~\bibnamefont
  {Wertz}}, \bibinfo {author} {\bibfnamefont {L.}~\bibnamefont {Ferrier}},
  \bibinfo {author} {\bibfnamefont {D.~D.}\ \bibnamefont {Solnyshkov}},
  \bibinfo {author} {\bibfnamefont {P.}~\bibnamefont {Senellart}}, \bibinfo
  {author} {\bibfnamefont {D.}~\bibnamefont {Bajoni}}, \bibinfo {author}
  {\bibfnamefont {A.}~\bibnamefont {Miard}}, \bibinfo {author} {\bibfnamefont
  {A.}~\bibnamefont {Lemaitre}}, \bibinfo {author} {\bibfnamefont
  {G.}~\bibnamefont {Malpuech}}, \ and\ \bibinfo {author} {\bibfnamefont
  {J.}~\bibnamefont {Bloch}},\ }\href {\doibase 10.1063/1.3192408} {\bibfield
  {journal} {\bibinfo  {journal} {Appl. Phys. Lett.}\ }\textbf {\bibinfo
  {volume} {95}},\ \bibinfo {pages} {051108} (\bibinfo {year}
  {2009})}\BibitemShut {NoStop}%
\bibitem [{\citenamefont {del Valle}\ \emph {et~al.}(2009)\citenamefont {del
  Valle}, \citenamefont {Sanvitto}, \citenamefont {Amo}, \citenamefont
  {Laussy}, \citenamefont {Andr\'{e}}, \citenamefont {Tejedor},\ and\
  \citenamefont {Vi\~{n}a}}]{DelValle2009}%
  \BibitemOpen
  \bibfield  {author} {\bibinfo {author} {\bibfnamefont {E.}~\bibnamefont {del
  Valle}}, \bibinfo {author} {\bibfnamefont {D.}~\bibnamefont {Sanvitto}},
  \bibinfo {author} {\bibfnamefont {A.}~\bibnamefont {Amo}}, \bibinfo {author}
  {\bibfnamefont {F.~P.}\ \bibnamefont {Laussy}}, \bibinfo {author}
  {\bibfnamefont {R.}~\bibnamefont {Andr\'{e}}}, \bibinfo {author}
  {\bibfnamefont {C.}~\bibnamefont {Tejedor}}, \ and\ \bibinfo {author}
  {\bibfnamefont {L.}~\bibnamefont {Vi\~{n}a}},\ }\href {\doibase
  10.1103/PhysRevLett.103.096404} {\bibfield  {journal} {\bibinfo  {journal}
  {Phys. Rev. Lett.}\ }\textbf {\bibinfo {volume} {103}},\ \bibinfo {pages}
  {096404} (\bibinfo {year} {2009})}\BibitemShut {NoStop}%
\bibitem [{\citenamefont {Christopoulos}\ \emph {et~al.}(2007)\citenamefont
  {Christopoulos}, \citenamefont {\relax{Baldassarri H\"{o}ger}~von
  H\"{o}gersthal}, \citenamefont {Grundy}, \citenamefont {Lagoudakis},
  \citenamefont {Kavokin}, \citenamefont {Baumberg}, \citenamefont
  {Christmann}, \citenamefont {Butt\'{e}}, \citenamefont {Feltin},
  \citenamefont {Carlin},\ and\ \citenamefont {Grandjean}}]{Christopoulos2007}%
  \BibitemOpen
  \bibfield  {author} {\bibinfo {author} {\bibfnamefont {S.}~\bibnamefont
  {Christopoulos}}, \bibinfo {author} {\bibfnamefont {G.}~\bibnamefont
  {\relax{Baldassarri H\"{o}ger}~von H\"{o}gersthal}}, \bibinfo {author}
  {\bibfnamefont {A.~J.~D.}\ \bibnamefont {Grundy}}, \bibinfo {author}
  {\bibfnamefont {P.~G.}\ \bibnamefont {Lagoudakis}}, \bibinfo {author}
  {\bibfnamefont {A.~V.}\ \bibnamefont {Kavokin}}, \bibinfo {author}
  {\bibfnamefont {J.~J.}\ \bibnamefont {Baumberg}}, \bibinfo {author}
  {\bibfnamefont {G.}~\bibnamefont {Christmann}}, \bibinfo {author}
  {\bibfnamefont {R.}~\bibnamefont {Butt\'{e}}}, \bibinfo {author}
  {\bibfnamefont {E.}~\bibnamefont {Feltin}}, \bibinfo {author} {\bibfnamefont
  {J.-F.}\ \bibnamefont {Carlin}}, \ and\ \bibinfo {author} {\bibfnamefont
  {N.}~\bibnamefont {Grandjean}},\ }\href {\doibase
  10.1103/PhysRevLett.98.126405} {\bibfield  {journal} {\bibinfo  {journal}
  {Phys. Rev. Lett.}\ }\textbf {\bibinfo {volume} {98}},\ \bibinfo {pages}
  {126405} (\bibinfo {year} {2007})}\BibitemShut {NoStop}%
\bibitem [{\citenamefont {Li}\ \emph {et~al.}(2013)\citenamefont {Li},
  \citenamefont {Orosz}, \citenamefont {Kamoun}, \citenamefont {Bouchoule},
  \citenamefont {Brimont}, \citenamefont {Disseix}, \citenamefont {Guillet},
  \citenamefont {Lafosse}, \citenamefont {Leroux}, \citenamefont {Leymarie},
  \citenamefont {Mexis}, \citenamefont {Mihailovic}, \citenamefont
  {Patriarche}, \citenamefont {R\'{e}veret}, \citenamefont {Solnyshkov},
  \citenamefont {Zuniga-Perez},\ and\ \citenamefont {Malpuech}}]{Li2013}%
  \BibitemOpen
  \bibfield  {author} {\bibinfo {author} {\bibfnamefont {F.}~\bibnamefont
  {Li}}, \bibinfo {author} {\bibfnamefont {L.}~\bibnamefont {Orosz}}, \bibinfo
  {author} {\bibfnamefont {O.}~\bibnamefont {Kamoun}}, \bibinfo {author}
  {\bibfnamefont {S.}~\bibnamefont {Bouchoule}}, \bibinfo {author}
  {\bibfnamefont {C.}~\bibnamefont {Brimont}}, \bibinfo {author} {\bibfnamefont
  {P.}~\bibnamefont {Disseix}}, \bibinfo {author} {\bibfnamefont
  {T.}~\bibnamefont {Guillet}}, \bibinfo {author} {\bibfnamefont
  {X.}~\bibnamefont {Lafosse}}, \bibinfo {author} {\bibfnamefont
  {M.}~\bibnamefont {Leroux}}, \bibinfo {author} {\bibfnamefont
  {J.}~\bibnamefont {Leymarie}}, \bibinfo {author} {\bibfnamefont
  {M.}~\bibnamefont {Mexis}}, \bibinfo {author} {\bibfnamefont
  {M.}~\bibnamefont {Mihailovic}}, \bibinfo {author} {\bibfnamefont
  {G.}~\bibnamefont {Patriarche}}, \bibinfo {author} {\bibfnamefont
  {F.}~\bibnamefont {R\'{e}veret}}, \bibinfo {author} {\bibfnamefont
  {D.}~\bibnamefont {Solnyshkov}}, \bibinfo {author} {\bibfnamefont
  {J.}~\bibnamefont {Zuniga-Perez}}, \ and\ \bibinfo {author} {\bibfnamefont
  {G.}~\bibnamefont {Malpuech}},\ }\href {\doibase
  10.1103/PhysRevLett.110.196406} {\bibfield  {journal} {\bibinfo  {journal}
  {Phys. Rev. Lett.}\ }\textbf {\bibinfo {volume} {110}},\ \bibinfo {pages}
  {196406} (\bibinfo {year} {2013})}\BibitemShut {NoStop}%
\bibitem [{\citenamefont {Tartakovskii}\ \emph {et~al.}(2000)\citenamefont
  {Tartakovskii}, \citenamefont {Emam-Ismail}, \citenamefont {Stevenson},
  \citenamefont {Skolnick}, \citenamefont {Astratov}, \citenamefont
  {Whittaker}, \citenamefont {Baumberg},\ and\ \citenamefont
  {Roberts}}]{Tartakovskii2000}%
  \BibitemOpen
  \bibfield  {author} {\bibinfo {author} {\bibfnamefont {A.~I.}\ \bibnamefont
  {Tartakovskii}}, \bibinfo {author} {\bibfnamefont {M.}~\bibnamefont
  {Emam-Ismail}}, \bibinfo {author} {\bibfnamefont {R.~M.}\ \bibnamefont
  {Stevenson}}, \bibinfo {author} {\bibfnamefont {M.~S.}\ \bibnamefont
  {Skolnick}}, \bibinfo {author} {\bibfnamefont {V.~N.}\ \bibnamefont
  {Astratov}}, \bibinfo {author} {\bibfnamefont {D.~M.}\ \bibnamefont
  {Whittaker}}, \bibinfo {author} {\bibfnamefont {J.~J.}\ \bibnamefont
  {Baumberg}}, \ and\ \bibinfo {author} {\bibfnamefont {J.~S.}\ \bibnamefont
  {Roberts}},\ }\href {\doibase 10.1103/PhysRevB.62.R2283} {\bibfield
  {journal} {\bibinfo  {journal} {Phys. Rev. B}\ }\textbf {\bibinfo {volume}
  {62}},\ \bibinfo {pages} {R2283} (\bibinfo {year} {2000})}\BibitemShut
  {NoStop}%
\bibitem [{\citenamefont {Qarry}\ \emph {et~al.}(2003)\citenamefont {Qarry},
  \citenamefont {Ramon}, \citenamefont {Rapaport}, \citenamefont {Cohen},
  \citenamefont {Ron}, \citenamefont {Mann}, \citenamefont {Linder},\ and\
  \citenamefont {Pfeiffer}}]{Qarry2003}%
  \BibitemOpen
  \bibfield  {author} {\bibinfo {author} {\bibfnamefont {A.}~\bibnamefont
  {Qarry}}, \bibinfo {author} {\bibfnamefont {G.}~\bibnamefont {Ramon}},
  \bibinfo {author} {\bibfnamefont {R.}~\bibnamefont {Rapaport}}, \bibinfo
  {author} {\bibfnamefont {E.}~\bibnamefont {Cohen}}, \bibinfo {author}
  {\bibfnamefont {A.}~\bibnamefont {Ron}}, \bibinfo {author} {\bibfnamefont
  {A.}~\bibnamefont {Mann}}, \bibinfo {author} {\bibfnamefont {E.}~\bibnamefont
  {Linder}}, \ and\ \bibinfo {author} {\bibfnamefont {L.~N.}\ \bibnamefont
  {Pfeiffer}},\ }\href {\doibase 10.1103/PhysRevB.67.115320} {\bibfield
  {journal} {\bibinfo  {journal} {Phys. Rev. B}\ }\textbf {\bibinfo {volume}
  {67}},\ \bibinfo {pages} {115320} (\bibinfo {year} {2003})}\BibitemShut
  {NoStop}%
\bibitem [{\citenamefont {Tartakovskii}\ \emph {et~al.}(2003)\citenamefont
  {Tartakovskii}, \citenamefont {Krizhanovskii}, \citenamefont {Malpuech},
  \citenamefont {Emam-Ismail}, \citenamefont {Chernenko}, \citenamefont
  {Kavokin}, \citenamefont {Kulakovskii}, \citenamefont {Skolnick},\ and\
  \citenamefont {Roberts}}]{Tartakovskii2003}%
  \BibitemOpen
  \bibfield  {author} {\bibinfo {author} {\bibfnamefont {A.~I.}\ \bibnamefont
  {Tartakovskii}}, \bibinfo {author} {\bibfnamefont {D.~N.}\ \bibnamefont
  {Krizhanovskii}}, \bibinfo {author} {\bibfnamefont {G.}~\bibnamefont
  {Malpuech}}, \bibinfo {author} {\bibfnamefont {M.}~\bibnamefont
  {Emam-Ismail}}, \bibinfo {author} {\bibfnamefont {A.~V.}\ \bibnamefont
  {Chernenko}}, \bibinfo {author} {\bibfnamefont {A.~V.}\ \bibnamefont
  {Kavokin}}, \bibinfo {author} {\bibfnamefont {V.~D.}\ \bibnamefont
  {Kulakovskii}}, \bibinfo {author} {\bibfnamefont {M.~S.}\ \bibnamefont
  {Skolnick}}, \ and\ \bibinfo {author} {\bibfnamefont {J.~S.}\ \bibnamefont
  {Roberts}},\ }\href {\doibase 10.1103/PhysRevB.67.165302} {\bibfield
  {journal} {\bibinfo  {journal} {Phys. Rev. B}\ }\textbf {\bibinfo {volume}
  {67}},\ \bibinfo {pages} {165302} (\bibinfo {year} {2003})}\BibitemShut
  {NoStop}%
\end{thebibliography}
\end{document}